 \documentclass[prd,nofootinbib,floats,aps,twocolumn,tightenlines,superscriptaddress,floatfix]{revtex4-1}

\usepackage{ifxetex}
\ifxetex
   \usepackage{polyglossia}
   \setdefaultlanguage{english}
   \usepackage{fontspec}
\else

   \usepackage[utf8]{inputenc}
\fi
\usepackage{graphicx,color}	
\usepackage[usenames,dvipsnames]{xcolor}
\usepackage{bbold}
 \usepackage{bm}
 
\usepackage{latexsym}
\usepackage{amsmath}
\usepackage{amsthm}
\usepackage{amsfonts}
\usepackage{amssymb}
\usepackage{textcomp}
\usepackage{cancel}
\usepackage{verbatim}
 \usepackage{hyperref}
\usepackage{enumerate}
\usepackage{cancel}
\usepackage{soul}

\allowdisplaybreaks

\newcommand{\be}{\begin{equation}}
\newcommand{\ee}{\end{equation}}

\newcommand{\ii}{\mathrm{i}}
\renewcommand{\vec}[1]{\bm #1}
\newcommand{\ket}[1]{\left| {#1} \right\rangle}
\newcommand{\bra}[1]{\left\langle {#1} \right|}

\def\ra{\rangle}

\makeatletter \def\cleardoublepage{\clearpage\if@twoside \ifodd\c@page\else
   \thispagestyle{empty}
   \newpage
   \if@twocolumn\hbox{}\newpage\fi\fi\fi}
\makeatother

\def\slashchar#1{\setbox0=\hbox{$#1$} 
\dimen0=\wd0 
\setbox1=\hbox{/} \dimen1=\wd1 
\ifdim\dimen0>\dimen1 
\rlap{\hbox to \dimen0{\hfil/\hfil}} 
#1 
\else 
\rlap{\hbox to \dimen1{\hfil$#1$\hfil}} 
/ 
\fi}

\renewcommand{\d}{\text{d}}

\newcommand{\cnm}[2]{\big[{#1},{#2}\big]}

\begin{abstract}
We study the transmission of information and correlations through quantum fields in cosmological backgrounds. With this aim, we make use of quantum information tools to quantify the classical and quantum correlations induced by a quantum massless scalar field in two particle detectors, one located in the early universe (Alice's) and the other located at a later time (Bob's). In particular, we focus on two phenomena: a)  the consequences  on the transmission of information  of the violations of the strong Huygens principle for quantum fields, and b) the analysis of the field vacuum correlations via correlation  harvesting from Alice to Bob. We will study  a  standard cosmological model first and then assess whether these results also hold if we use other than  the  general relativistic dynamics.  As a particular example,  we will study the  transmission of information  through the Big Bounce,  that replaces the Big Bang, in  the  effective dynamics  of Loop Quantum Cosmology.

 \end{abstract}

\begin{document}

\title{  Timelike information broadcasting in cosmology}
\author{Ana Blasco}
\affiliation{Departamento de F\'{\i}sica Te\'orica II, Universidad Complutense 
de Madrid, 28040 Madrid, Spain}
\author{Luis J. Garay}
\affiliation{Departamento de F\'{\i}sica Te\'orica II, Universidad Complutense 
de Madrid, 28040 Madrid, Spain}
\affiliation{Instituto de Estructura de la Materia (IEM-CSIC), Serrano 121, 28006 Madrid, Spain}
\author{Mercedes Mart\'{\i}n-Benito}
\affiliation{Radboud University Nijmegen, Institute for Mathematics, Astrophysics and Particle Physics, Heyendaalseweg 135, NL-6525 AJ Nijmegen, The Netherlands}
\author{Eduardo Mart\'{\i}n-Mart\'{\i}nez }
\affiliation{Institute for Quantum Computing, University of Waterloo, Waterloo, Ontario, N2L 3G1, Canada}
\affiliation{Department of Applied Mathematics, University of Waterloo, Waterloo, Ontario, N2L 3G1, Canada}
\affiliation{Perimeter Institute for Theoretical Physics, Waterloo, Ontario, N2L 6B9, Canada}

\maketitle

\setcounter{secnumdepth}{3} 
\setcounter{tocdepth}{3}    


\section{Introduction}

The quest  for gathering knowledge about the  very early universe and its evolution is one of the most challenging endeavors in physics. Quantum field theory on curved spacetime  has, in many cases, offered a suitable framework  for the exploration of quantum phenomena in cosmology. The most prominent example of its application is our current cosmological paradigm, based on the theory of cosmological perturbations supplemented with inflation  (see e.g. \cite{Langlois:2010xc,Mukhanov:2005sc,liddle2000cosmological,Martin:2004um}). This framework remarkably succeeds in explaining the  large scale structure of the Universe. Cosmological perturbations are treated as quantum vacuum fluctuations in the early universe. These fluctuations turned into classical density anisotropies that left an imprint in the cosmic microwave background \cite{Lahav:2014vza,Ade:2015xua}. They were the seeds  giving rise to the galaxies and other structures that we observe nowadays. 

The analysis of quantum vacuum fluctuations is not limited to the scenario above. Actually, vacuum fluctuations and  quantum entanglement  are at the core of a plethora of phenomena such as the Unruh effect \cite{Unruh1976}, Hawking radiation \cite{Hawking1975}, and the Gibbons-Hawking effect \cite{GibHawking}. In the context of cosmology, the role played by vacuum entanglement has been recently reviewed in \cite{cosmoq,review2}. Moreover,  it is known that  vacuum entanglement  could in principle  be exploited to detect spacetime curvature  \cite{Steeg}, or as a powerful physical resource to encode and transmit  classical and quantum information, as it has been explored in recent years \cite{reznik,Olson2011,Collapse2,Comm1,Jonsson:2014lja}.  A natural question then arises: could it be possible to  make use of these correlations to gain access to information  about  early universe events?  If this were the case, they could have left observable imprints on the fine details of the cosmic microwave background. And furthermore,  how could we extract this information broadcast through the universe? The present  work aims to deepen our understanding  of  these questions, in particular     the transmission of information from earlier stages of the universe to later epochs,  by further exploiting  the analysis of quantum and classical correlations of quantum fields in the context of cosmology. We will first consider the dynamics provided by the theory of general relativity, which for many matter contents predicts a Big Bang.  Then, as an example to study transmission of information between two branches of a bouncing universe, we will adopt the effective dynamics derived from Loop Quantum Cosmology (LQC), that replaces the classical Big Bang singularity with a Big Bounce  \cite{Ashtekar:2006wn,Bojowald:2008zzb,Banerjee:2011qu,Ashtekar:2011ni}. 

More precisely, we will consider spatially flat Friedmann-Robertson-Walker (FRW) expanding cosmologies  generated by a perfect fluid, and a test massless scalar field  coupled to the background geometry, that we will quantize in the adiabatic vacuum.  In this context, observers comoving with the Hubble flow, and probing the field by locally coupling detectors, will naturally perceive  particle production due to the spacetime expansion  \cite{GibHawking}. 
We will introduce a pair of those comoving observers: Alice, living in an early stage of the universe, and  Bob, living in a latter epoch, each of them will be equipped with a detector interacting with the field.

Owing to their interaction with the field, Alice's and Bob's detectors not only carry information about the dynamics of the universe \cite{Garay:2013dya}, but also may become entangled, since the vacuum state of the field is entangled. This swapping of entanglement between the field and the detectors was first studied in \cite{Valentini1991321,reznik}, where it was shown that field entanglement can  be extracted by local quantum systems interacting with the field even if they are spacelike separated. This phenomenon, which also receives the name of {\it entanglement harvesting} \cite{cosmoq,Nicklast}, has been proposed as a sustainable resource for quantum information via the so-called quantum field {\it entanglement farming} protocols \cite{farming}. By using quantum information tools, we will quantify the correlations shared by the early-universe observer Alice and the late-universe observer Bob for the case in which the field is conformally coupled to the geometry. This analysis will reveal this correlation-harvesting phenomenon, showing that it is non-vanishing even when Alice and Bob are not causally connected.
  
The presence of correlations in the final state of the detectors will be an indication  not only of  vacuum entanglement \cite{reznik} but also  of the exchange of field quanta \cite{Comm1}. When Alice and Bob are causally connected, Alice's detector coupling to the field provokes  perturbations in the vacuum that will eventually reach Bob. These correlations in  the states of the detectors suggest  the possibility of establishing a communication channel in cosmological timescales.
This question is closely connected with recent results in relativistic quantum communication
\cite{Comm1,Jonsson:2014lja}. 
In (four-dimensional) flat spacetime,   and for massless fields, communication can only occur at the speed of light. However in lower dimensions (and also in higher odd dimensions), or in the presence of curvature, signals leak into the timelike area of the light cone. This phenomenon  stems from the violation of the strong Huygens principle  \cite{Ellis,McLenaghan,Sonego:1991sq,czapor}, which in the study of propagation of  classical   waves is also known as the `tails problem' (see e.g. \cite{Blanchet:1987wq,Blanchet:1992,Bombelli:1994rh,Pullin,Hod:1999ci} for references in the context of gravity). 
In quantum field theory,  the violations of the strong Huygens principle allow for slower-than-light communication using massless fields, by means of protocols where information can be communicated without transmitting energy from Alice to Bob and where the energy cost of sending a message is spent by the receiver of the message  in the action of reading it out \cite{Jonsson:2014lja}.  
These violations have also been  studied before in the context of cosmology for classical fields \cite{Faraoni:1991xe, Faraoni:1999us}.

In this work, we analyze the consequences of the violations of the strong Huygens principle in the transmission of quantum information in cosmology.  In \cite{Blasco:2015eya}, we already presented part of this study. More precisely, we analyzed the capacity of the communication channel enabled by the violations of the strong Huygens principle in a cosmological model. This analysis was carried out both for minimal and conformal coupling between the field and the curvature. We showed that, while for the former there is no violation of the strong Huygens principle due to conformal invariance, for the later there is violation  and that this implies that not only lightlike but also that timelike communication is possible. In the present work, we present the details of that previous  analysis and extend  it in  three  directions: i) we particularize the study not only to a matter-dominated universe, as in \cite{Blasco:2015eya}, but also to a universe generated by a cosmological constant;  ii) we consider arbitrary detector gaps  and present the  zero gap  case considered in \cite{Blasco:2015eya} as a particular case; and iii) 
we deal not only with the general relativistic dynamics, as in \cite{Blasco:2015eya}, but also with the LQC bouncing dynamics.   
Even though the channel capacity turns out to decay with the temporal separation of the detectors, we will show that it is possible to compensate this decay by  including additional receivers. Remarkably, for timelike communication, the channel capacity  is independent of the spatial separation of the detectors. These results open a new door to an exciting set of resources  to  have access to information coming from the early phases of our universe that  have been  so far unexplored.

The outline of this article is as follows:  Sec. \ref{sec:setting}  introduces  the basic setting, including the description of the background geometry, the way we model the detectors-field interaction, and the computation of the evolved states of the detectors.   Sec. \ref{sec:signaling} develops the theoretical framework needed to estimate the capacity of the communication channel established between Alice and Bob. Making use of those results, Sec. \ref{sec:GR-sig} is devoted to the study of the transmission of information between Alice and Bob, and the violations of the strong Huygens principle, in the standard cosmological model. We will see in Sec. \ref{sec:LQC-sig} that  the observed phenomenology is also present in the case of the effective LQC dynamics. In Sec. \ref{sec:mutual} we will analyze  the correlations harvesting phenomenon, by computing the mutual information.  
 Finally, Sec. \ref{sec:con} will be devoted to the  conclusions.  Natural units $\hbar = c = 1$ are used throughout.

\section{Dynamical framework}
\label{sec:setting}

\subsection{Background geometry and test field}
\label{sec:geometry}

For the spacetime geometry we choose the simplest nontrivial cosmology, namely an open and spatially flat FRW spacetime. The corresponding metric is conformally flat,
\be \label{eq:FRW}
\d s^2=a(\eta)^2(-\d \eta^2+\d r^2+r^2\d\Omega^2),\ee
where  $a$ is the scale factor. Here $\eta$ is the conformal time, $r$ a radial coordinate, and  $\d\Omega^2$ the metric in the unit 2-sphere. 
The coordinates of comoving observers are given by $t$, $r$, and the solid angle $\Omega$. The comoving time $t$ is related to the conformal time via
$\d t/ \d \eta=a$. We consider that this geometry is generated by a perfect fluid with a constant pressure-to-density ratio,
$p/ \rho=w\geq-1$.

The  general relativistic  equations of motion for the scale factor and the energy density (the Friedmann  and the continuity equations) are given by 
\begin{align}
\label{class}
\left(\frac1{a}\frac{\d a}{\d t}\right)^2=\frac{8\pi G}{3}\rho, \qquad \rho\propto a^{-3(w+1)}.
\end{align}
For $w>-1$, expanding solutions  evolve as 
\be \label{class-sol} a \propto t^{\frac{2 }{3w+3}}\propto\eta^{ \frac{2}{3w+1}},\ee 
with $t, \eta\in [0,\infty)$.
For $w=-1$, the universe is generated by a cosmological constant $\Lambda$, with $|\Lambda|=8\pi G\rho/3$. In this case expanding solutions evolve as
\be a \propto e^{\sqrt{|\Lambda|} t}  \propto - \frac{1}{\sqrt{|\Lambda|}\eta},\ee with $t\in(-\infty,\infty)$ and $\eta\in(-\infty,0)$.

According to general relativity, a universe dominated by a perfect fluid with $w>-1$ arose from a Big Bang singularity, as the matter energy density $\rho$, or equivalently the curvature $\mathcal R$, diverges at initial time $t=0$. However, we do not have reasons to trust general relativity in regimes where the energy densities become Planckian, as in those regimes quantum effects of the geometry might become important.
Indeed, the dynamics predicted by general relativity changes drastically if we allow for modifications of quantum geometric nature. That is for instance the case if we consider LQC  \cite{Bojowald:2008zzb,Banerjee:2011qu,Ashtekar:2011ni}.

LQC is a background-independent quantization for cosmological spacetimes that adopts a so-called polymeric representation for the geometric degrees of freedom. This quantum representation renders the microscopic structure of the geometry discrete. As a consequence, for classical models displaying a Big Bang, the quantum evolution replaces the singularity by a quantum bounce, where physical observables do not diverge. For semiclassical states, the Planck regime serves as a bridge between two large classical universes: a contracting cosmological phase and an expanding one become deterministically connected via a Big Bounce  \cite{Ashtekar:2006wn}.  

This bounce scenario opens the possibility of analyzing the transmission of information from the contracting branch of the universe to the expanding one. It is then an interesting scenario for the communication protocols  explored in \cite{Jonsson:2014lja, Blasco:2015eya} and we will analyze it in this paper. For this we also need to introduce the spacetime dynamics in the LQC scenario.

In the loop quantization, the observable representing the scale factor, when computed on appropriate semiclassical states, displays expectation values along a smooth trajectory and negligible relative fluctuations. 
It is therefore possible to derive an effective dynamics for these states \cite{Taveras:2008ke,Ashtekar:2006wn}, which leads to the following modified Friedmann equation:
\begin{align}\label{eff}
\left(\frac{\dot a}{a}\right)^2=\frac{8\pi G}{3}\rho\Big( 1-\frac{\rho}{\rho_\star}\Big).
\end{align}
Here, $\rho_\star=6\pi G/l^6$ is the critical energy density (maximum eigenvalue of the density operator in the loop quantization \cite{Ashtekar:2007em});
$l^3$ is a parameter of the quantization that gives essentially the quanta of volume (in LQC the volume has a discrete spectrum equally spaced by $2l^3$ units) and depends on other fundamental parameters of loop quantum gravity \cite{Banerjee:2011qu,Ashtekar:2011ni}). 
This dynamics departs from that of general relativity when $\rho$ reaches the critical density $\rho_\star$, turning the classical Big Bang into a Big Bounce. General relativity is recovered in the limit $\rho_\star\rightarrow\infty$, or equivalently $l\rightarrow 0$. 

The continuity equation for the energy density $\rho$ does not change, and therefore the solution to  \eqref{eff}  for $w>-1$  is
\be 
\label{eq:scale_LQC} a(t)=  \left\{ 6\pi G \beta \left[(w+1) t\right]^2+\frac{\beta}{\rho_\star}\right\}^\frac1{3(w+1)},
\ee
where $\beta=\rho a^{3(w+1)}$ is constant. 
In turn, the explicit relation between conformal and comoving times becomes
\begin{align} \label{eq:eta_LQC} 
\eta ( t)&=\mathcal{C}\; {}_2F_1 \left[\frac{1}{3(w+1)},\frac{1}{2};\frac{3}{2};-6\pi G \rho_\star\left[(w+1) t\right]^2\right] t  ,
\end{align}
where $_2F_1$ is an ordinary hypergeometric function and  we have defined  the constant
$\mathcal{C}=\left( \rho_\star/\beta\right)^{\frac1{3(w+1)}}$.  
Note that, unlike in the classical case where we have either the expanding solution of Eq. \eqref{class-sol} with $t\in[0,\infty)$, or its time reversal with $t\in(-\infty,0]$, now $t\in(-\infty,\infty)$ and a bounce deterministically joins these two branches of the universe. At the bounce point, $t=0$, the scale factor does not vanish and there is no singularity. Notice that for the cosmological constant case, the general relativistic dynamics and the LQC effective dynamics completely agree, as this case does not develop a classical curvature singularity. This equivalence makes redundant the analysis of the case $w=-1$ in the study of the LQC effective dynamics.

In these spacetime backgrounds (for any $w\ge -1$)  we will introduce a test massless scalar field $\phi$ quantized in the adiabatic vacuum \cite{Birrell}. As it has been shown in \cite{guille0,guille2,guille1}, for conformally flat compact spacetimes  there exist natural criteria that select a unique equivalence class of vacua, which includes the adiabatic vacuum. These criteria are invariance of the vacuum under the spatial isometries and unitarity of the vacuum dynamics. Extrapolating these results to our setting we select as the initial field state the adiabatic vacuum. This is also convenient because the adiabatic vacuum is the initial field state for which the creation of particles due to the expansion of the spacetime is finite and the smallest possible.  Notice, however, that as discussed in \cite{Jonsson:2014lja,Blasco:2015eya,Martin-Martinez:2015} and as we will show explicitly later, the results of sections  \ref{sec:signaling},  \ref{sec:GR-sig} and \ref{sec:LQC-sig}  are independent of the initial state of the field. The results of section  \ref{sec:mutual}  depend on the initial state but would only change quantitatively and not qualitatively if the initial state of the field is altered. 

The equation of motion for the test field $\phi$ is
\begin{equation}\label{eq:fieldeq}
\left(\Box-\xi \mathcal{R} \right)\phi=0,
\end{equation}
where $\xi$ is the coupling of the field to the Ricci scalar
\be\label{R}
\mathcal{R}=\frac{6}{a^{3}} \frac{\d^2 a}{ \d\eta^2},
\ee
and the d'Alembertian operator acquires the form
\be \label{box}\Box=-\frac1{a^4}\frac{\d}{\d\eta}\left(a^2\frac{\d}{\d\eta}\right)+\frac1{a^2}   \nabla^2,\ee
being $\nabla^2$ the Laplacian in $\mathbb{R}^3$.

Note that since the field is prepared in the adiabatic vacuum, the expectation value of its stress-energy tensor, and in particular the energy density $\langle T_{tt}\rangle$, can be considered very small as compared with the energy density of the fluid that generates the curvature, namely $\rho\gg \langle T_{tt}\rangle$. Therefore we neglect the backreaction of the field on the gravitational background. 

As anticipated in the Introduction, when analyzing the signaling and the violations of the strong Huygens principle in Sec. \ref{sec:signaling}, we will consider minimal coupling ($\xi=0$)  and conformal coupling ($\xi=1/6$ for the 4-dimensional case) of the massless field to the Ricci curvature. However, for the analysis of correlations of Sec. \ref{sec:mutual}, we will only consider conformal coupling. The reason is that, as we will discuss below, the conformal coupling scenario is devoid of violations of the strong Huygens principle, and thus the contributions to the detector correlations harvested from the field will not be masked by timelike signaling, allowing us to  identify vacuum correlation harvesting  (classical and quantum, but not necessarily entanglement \cite{Pozas-Kerstjens:2015})   in a cleaner way.

\subsection{Detectors-field interaction}

In a homogeneous and isotropic universe, as the one that we are considering, there exists a family of privileged observers, called comoving observers, who perceive an isotropic space-time evolution because they move along the Hubble flow. The proper time of comoving observers does not coincide with the conformal time. Let us recall that the conformal time is the natural parameter associated with the conformal timelike Killing vector of this geometry, hence it is used to define the adiabatic vacuum. Since conformal time is not proper to comoving observers, a comoving observer will detect particles even if the usual particle creation  associated with curvature is tamed by conformal invariance, as it is the case for the conformal vacuum in the conformal coupling case \cite{cosmoq}. This phenomenon is the well-known Gibbons-Hawking effect \cite{GibHawking}. So in this sense, the spacetime expansion or contraction leads to a particle production as seen by observers comoving with the Hubble flow. 

We now introduce a couple of observers Alice  and Bob.  Alice lives in the very early-universe while Bob lives  at a later time in the expanding classical universe. 
We consider Alice and Bob to be comoving observers in the adiabatic vacuum, and thus both of them will naturally detect particles. They do not have direct access to the field, but they can perform measurements on it indirectly by locally coupling `particle detectors' to the field.

We will model these particle detectors  by a pair of  two-level quantum systems (qubits). The subindex \mbox{$\nu=\{A,B\}$} will be used to denote either Alice's or Bob's detector.  Let us define the  interaction-picture   monopole moment of the detector $\nu$  as
\be \label{eq:monop}
 \mu_\nu(t)= \sigma^+_\nu e^{\ii\Omega_{\nu} t} +\sigma^-_\nu e^{-\ii\Omega_{\nu} t},\ee 
where we have used the following standard notation: \mbox{$\sigma_\nu^+=|e_\nu\rangle\langle g_\nu|$} and $\sigma_\nu^-=|g_\nu\rangle\langle e_\nu|$, for the $SU(2)$ ladder operators, $|g_\nu\rangle$ and $|e_\nu\rangle$ are the detector's ground and excited states, and $\Omega_\nu$ is its energy gap.
Moreover, we will consider that these detectors are spatially smeared with a  spherical  Gaussian distribution \be\label{eq:smeared} 
F(\bm x,t)=\frac{1}{\sigma^3 \sqrt{\pi^3}}e^{-\frac{a(t)^2}{\sigma^2 } \vec{x}^2}.
\ee
Here $\sigma$ characterizes the physical size of the detectors.  The smearing is specifically introduced to regularize the UV divergences that can appear in the case of point like detectors \cite{Wavepackets}, apart from the obvious fact that any realistic particle detector has a finite size.

The interaction between the field and the particle detectors will be described by the  Unruh-DeWitt  model \cite{DeWitt}. Although simple, this interaction model, widely used in the  literature, already displays all the fundamental features of the light-matter interaction when there is no exchange of orbital angular momentum \cite{Alvaro,Wavepackets}. The Unruh-DeWitt interaction Hamiltonian (in the interaction picture) for each detector is given by
\be \label{eq:HI}
  {H}_{I,\nu}=\lambda_\nu \chi_\nu(t)   \mu_\nu(t) \int \text{d}^3 \vec{x}\,a(t)^3 F[\bm x-\bm x_{\nu}(t),t]    \phi[\bm x,\eta(t)].\ee
Here $t$ is the proper time of both detectors, considered to be comoving; the spatial smearing $F[\bm x-\bm x_{\nu}(t),t] $
is centered at $\bm x_{\nu }(t)$, the detector's trajectory, which for the  comoving case becomes $\bm x_\nu=$ const; $0\le\chi_\nu(t)\le 1$ is the detector's switching function; and
$\lambda_\nu$  is the coupling strength. 
The field operator $\phi$ is evaluated along the detector's worldline  integrated over the whole spatial extension of the detector. 

We can expand the field operator in  terms of positive and negative frequency solutions of Eq. \eqref{eq:fieldeq}, $u_{\vec{k}}$ and $u^*_{\vec{k}}$,
\be
  {\phi}(\vec{x} ,t)=\int \d^3\vec{k}[  {a}_{\vec{k}} u_{\vec{k}}(\vec{x} ,\eta(t))+  {a}_{\vec{k}}^\dagger u^*_{\vec{k}}(\vec{x} ,\eta(t))],
\ee
with $  {a}_{\vec{k}}$ and ${  {a}_{\vec{k}}}^\dagger$ the usual annihilation and creation operators satisfying the commutation relations: $[  {a}_{\vec{k}},{  {a}_{{\vec{k}}'}}^\dagger]=\delta({\vec{k}}-{\vec{k}}')$ and $[  {a}_{\vec{k}},{  {a}_{{\vec{k}}'}} ]=0$. Introducing the detector's form factor via the Fourier transform of the spatial smearing
\be \label{Fou}
\tilde F( \bm k,t)=\int_{\mathbb{R}^3} \d^3 \vec{x}\, a(t)^3 F(\bm x,t)e^{\ii \bm k\cdot \bm x}=e^{-\frac{\sigma^2   \bm k^2}{4a(t)^2} },\ee
Eq. \eqref{eq:HI} can be written as 
\begin{align}\label{eq:HI2}
  {H}_{I,\nu}&=\lambda_\nu   \chi_\nu(t)   \mu_\nu(t)\int \d^3\vec{k}\,\tilde F( \bm k,t)\Big(  {a}_{\vec{k}} u_{\vec{k}}[\vec{x}_\nu ,\eta(t)]\nonumber\\
&+  {a}_{\vec{k}}^\dagger u^*_{\vec{k}}[\vec{x}_\nu ,\eta(t)]\Big).
\end{align}

For simplicity, we will consider that both detectors are suddenly switched on and off, so the switching functions are given by 
 \begin{align}\label{eq:switching}
\chi_{\nu}(t)&= \left\{ \begin{array}{ll}
1, \qquad&    t\in[T_{i\nu},T_{f\nu}],\\
0,&  t\not\in[T_{i\nu},T_{f\nu}].\\
\end{array}\right.
\end{align}
As anticipated,  the typical divergences present in the case of a point-like detector with sudden switching are avoided here due to the fact that our detectors have a finite spatial size \cite{Langlois:2005nf,Louko:2006zv,Louko:2006yf}.

\subsection{Evolved state of the detectors}

At initial time $T_0$, we consider the test field $\phi$ to be in the adiabatic vacuum state $|0\ra$ and both detectors to be in arbitrary uncorrelated states, $\rho_{0,A}$ and $\rho_{0,B}$. 
The initial density matrix of the coupled system detectors-field is  therefore
\begin{align}
\rho_0&=\rho_{0,A}\otimes\rho_{0,B}\otimes \ket 0 \bra0.
\end{align}
After some time $T$, the evolved density matrix  will be
   \be \label{eq:roT}
   \rho_T=U\rho_0 U^{\dag},
   \ee
where
\be
U=\mathcal{T}\exp\left[-\ii \int_{T_0}^T \!\!\d t\, H_I(t)\right] 
\ee
is the evolution operator generated by the time-dependent Hamiltonian $  {H}_I=  {H}_{I,A}+  {H}_{I,B}$ of the two detectors given in Eq. \eqref{eq:HI2}. Here $\mathcal{T}$ denotes time-ordered exponential.

We can compute  the density matrix $\rho_T$  using perturbation theory for the evolution operator $U$, as long as
the coupling strengths $\lambda_{\nu}$  are small enough. Up to second order in the perturbative expansion, we have
\be
U=\openone+U^{(1)}+U^{(2)}+O(\lambda_\nu^3),
\ee
where
   \begin{align}
U^{(1)}&=-\ii \int_{T_0}^T \d t_1   {H}_I(t_1),\\*
U^{(2)}&=-\int_{T_0}^T \d t_1\int_{T_0}^{t_1} \d t_2   {H}_I (t_1)  {H}_I (t_2).
 \end{align}
The perturbative expansion of the time-evolved density matrix then reads 
\be \label{eq:roT2}
\rho_T=\rho_0+\rho_{T}^{(1)}+\rho_{T}^{(2)}+O(\lambda_\nu^3),
\ee
with
\begin{align}\label{eq:orders}
\rho_{T}^{(1)}&=U^{(1)}\rho_0+\rho_0 U^{(1)\dag},\\
\label{eq:rho2}\rho_{T}^{(2)}&=U^{(2)}\rho_0+\rho_0 U^{(2) \dag}+U^{(1)}\rho_0 U^{(1) \dag}.\nonumber
   \end{align}

The partial density matrix of the sub-system $AB$ formed by the two detectors is obtained by tracing out the field degrees of freedom,
  \be \label{eq:tracefield}
  \rho_{T,AB}=\text{tr}_{\phi}(\rho_T)\simeq \text{tr}_{\phi}(\rho_0+\rho_{T}^{(1)}+\rho_{T}^{(2)})=\text{tr}_{\phi}(\rho_0+\rho_{T}^{(2)}).
  \ee
Let us note that $\text{tr}_{\phi}(\rho_T^{(1)})=0$ since $\rho_T^{(1)}$ only contains non-diagonal terms in the field. Hence, at first order in perturbation theory the partial density matrix $ \rho_{T,AB}$ does not evolve, and we need to go at least to second order.

 The terms in $\rho_{T}^{(2)}$ are given by
 \begin{align}
&U^{(1)}\rho_0 U^{(1)\dag}=\int_{T_0}^{T} \d t_1\int_{T_0}^{T} \d t_2   {H}_I(t_1)\rho_0   {H}_I (t_2),\nonumber \\
&U^{(2)}\rho_0=-\int_{T_0}^T \d t_1\int_{T_0}^{t_1} \d t_2    H_I (t_1)   H_I (t_2)\rho_0.
\end{align}

Once $\rho_{T,AB}$  has been obtained, to get Alice and Bob detectors' final density matrices, $\rho_{T,A}$ and $\rho_{T,B}$, we trace out  $B$ and $A$ respectively from \eqref{eq:tracefield}:
\begin{align}
 \label{eq:rhoA}
 \rho_{T,A}&=\text{tr}_B(\rho_{T,AB}),&
 \rho_{T,B}&=\text{tr}_A(\rho_{T,AB}).
 \end{align}
 
 \subsection{A particular example of time evolution: Conformal coupling}
 \label{rhos-conformal}

Notice that our study  will not be  limited to conformally coupled fields. In fact we will also pay special attention to the minimal coupling scenario since that is the setup that allows for the violations of the Strong Huygens principle, and then for timelike communication \cite{Blasco:2015eya}. However, for illustration, let us compute the explicit expressions of $\rho_{T,AB}$, $\rho_{T,A}$, and $\rho_{T,B}$ for the simple case of the conformally coupled field quantized in the conformal vacuum, that we will later use for the study of generalized correlations and mutual information. 

We write the density matrix of the system $AB$ in the form
 \be\label{eq:rhoAB}
  \rho_{T,AB}=\rho_{0,AB}+\rho^1_{AB}+\rho^2_{AB}+(\rho^2_{AB})^{\dagger},\ee
 where the superindices $1,2$ on the rhs of \eqref{eq:rhoAB} indicate that the terms originate from $U^{(1)}\rho_0 U^{(1) \dag}$ or $U^{(2)}\rho_0$ in \eqref{eq:orders} respectively. After some calculations,  we obtain the following explicit result,
\begin{align}\label{di}
\rho_{0,AB}&=\rho_{0,A}\otimes \rho_{0,B},\\
\rho^1_{AB}=\sum_{\epsilon,\delta}&\bigg[M_A^{\epsilon,\delta} (\sigma_A^\delta\rho_{0,A}\sigma_A^\epsilon)\otimes\rho_{0,B}\nonumber\\&+
M_{B,A}^{\epsilon,\delta} (\sigma_A^\delta\rho_{0,A})\otimes (\rho_{0,B}\sigma_B^\epsilon) \nonumber\\ 
 &+M_{A,B}^{\epsilon,\delta} (\rho_{0,A}\sigma_A^\epsilon)\otimes (\sigma_B^\delta\rho_{0,B}) \nonumber\\ & + M_B^{\epsilon,\delta} \rho_{0,A}\otimes(\sigma_B^\delta\rho_{0,B}\sigma_B^\epsilon)\bigg],\\
\rho^2_{AB}=-\sum_{\epsilon,\delta}&\bigg[N_{A}^{\epsilon,\delta}[\sigma_A^\epsilon\sigma_A^\delta\rho_{0,A}]\otimes\rho_{0,B}\nonumber\\
&+N_{B}^{\epsilon,\delta}\rho_{0,A}\otimes [\sigma_B^\epsilon\sigma_B^\delta\rho_{0,B}] \nonumber\\ & + M_{B,A}^{\epsilon,\delta} (\sigma_A^\delta\rho_{0,A})\otimes(\sigma_B^\epsilon\rho_{0,B})\bigg].
\end{align}
where we have made use of the fact that the support of $\chi_B(\eta)$ is in the strict future of the support of $\chi_A(\eta)$. In the expressions above we have introduced the following integrals:
\begin{align}\label{eq:I1}
M_\nu^{\epsilon,\delta}&=\frac{2\lambda _{\nu}^2 }{ (2\pi\sigma)^2} \nonumber\\*&\times \int_{\eta_{i\nu}}^{\eta_{f\nu}} \!\!\!\d \eta_1 e^{\ii\epsilon\Omega_\nu t(\eta_1)}\int_{\eta_{i\nu}}^{\eta_{f\nu}}  \!\!\!\!  \d \eta_2 e^{\ii\delta\Omega_\nu t(\eta_2)}\; J(\eta_1,\eta_2),\\
N_\nu^{\epsilon,\delta}&=\frac{2\lambda _{\nu}^2 }{ (2\pi\sigma)^2}  \nonumber\\&\times  \int_{\eta_{i\nu}}^{\eta_{f\nu}} \!\!\!\d \eta_1 e^{\ii\epsilon\Omega_\nu t(\eta_1)} \int_{\eta_{i\nu}}^{\eta_1}  \!\!\!\!  \d \eta_2 e^{\ii\delta\Omega_\nu t(\eta_2)} \; J(\eta_1,\eta_2),\\\label{eq:I3}
M^{\epsilon,\delta}_{\nu_1,\nu_2}&=\frac{\lambda _{\nu_1}\lambda _{\nu_2} }{ (2\pi)^2 R}  \frac{\sqrt{\pi}}{2\sigma}\nonumber\\
&\times \int_{\eta_{i\nu_1}}^{\eta_{f\nu_1}} \!\!\!\d \eta_1  e^{\ii\epsilon\Omega_\nu t(\eta_1)} \int_{\eta_{i\nu_2}}^{\eta_{f\nu_2}}  \!\!\!\!  \d \eta_2 e^{\ii\delta\Omega_\nu t(\eta_2)}\; K(\eta_1,\eta_2),
\end{align}
where $\epsilon$ and $\delta$ are either plus or minus; we have already used Eq. \eqref{eq:switching} to set the integration limits;  we have denoted $\eta_{i\nu}\equiv\eta(T_{i\nu})$, $\eta_{f\nu}\equiv\eta(T_{f\nu})$, and $R\equiv\|\vec{x}_A-\vec{x}_B\|$. The functions $J$ and $K$ in the integrands of Eqs. (\ref{eq:I1}--\ref{eq:I3}) are given by
   \begin{align}\label{eq:Jeq}
&J(\eta_1,\eta_2)
= b(\eta_1,\eta_2)+\ii \frac{\sqrt{\pi}}{\sigma}(\eta_2-\eta_1)b(\eta_1,\eta_2)^{3/2}\nonumber\\&\times e^{-\frac{(\eta_2-\eta_1)^2b(\eta_1,\eta_2)}{\sigma^2}}\Bigg[1 +\text{erf}\left(\ii \frac{(\eta_2-\eta_1)\sqrt{b(\eta_1,\eta_2)}}{\sigma}\right)\Bigg],\\
&K(\eta_1,\eta_2)= \ii\sqrt{b(\eta_1,\eta_2)}e^{-[f_+(\eta_1,\eta_2)]^2}\Bigg\{\text{erf}\left[\ii f_+(\eta_1,\eta_2)\right]\nonumber\\&-1+e^{\frac{4(\eta_2-\eta_1)Rb(\eta_1,\eta_2)}{\sigma^2}} \bigg[1-\text{erf}\left[\ii f_-(\eta_1,\eta_2)\right]\bigg]\Bigg\},
\end{align}
where
\begin{align}
b(\eta_1,\eta_2)&=\frac{a(\eta_1)^2 a(\eta_2)^2}{a(\eta_1)^2+a(\eta_2)^2},\\
f_\pm(\eta_1,\eta_2)&=\frac{(R\pm\eta_2\mp\eta_1)\sqrt{b(\eta_1,\eta_2)}}{\sigma},
\end{align}
and $\text{erf}(z)$ denotes the error function  \cite{abramovitz-stegun}.

We now have all the ingredients needed to compute the partial density matrices $\rho_{T,AB}$, $\rho_{T,A}$ and $\rho_{T,B}$.

 \section{ Signaling estimator and Channel Capacity} \label{sec:signaling}

We are going to quantify the amount of information that the early observer Alice sends to the later observer Bob using the quantum field and local interactions with it through particle detectors.
We will start by computing the signaling estimator $S$ defined in \cite{Jonsson:2014lja}. 
This estimator measures how the interaction of $A$ with the field influences the excitation probability of $B$. Let \mbox{$|\psi_{0,\nu}\rangle=\alpha_\nu|e_\nu\rangle+\beta_\nu|g_\nu\rangle$} be the initial state of the detector $\nu$. Adopting the basis
\be
|e_\nu\rangle=\left(
\begin{array}{c}
1\\
0\\
\end{array}
\right),\qquad |g_\nu\rangle=\left(
\begin{array}{c}
0\\
1\\
\end{array}
\right),
\ee
then the excitation probability of $B$ is given by the first component of the evolved partial density matrix $ \rho_{T,B}$. In consequence, the estimator $S$ is the contribution to that component that is proportional to $\lambda_A\lambda_B$, and it is given by
\be\label{S}
S=\lambda_A\lambda_B S_2+\mathcal O(\lambda_\nu^4),\ee where
 \begin{align}\label{signaling}
S_2\!&=\!4\!\int\! \d v\!\!  \int \!\d v' \chi_A(t)\chi_B(t') \text{Re}(\alpha^*_A\beta_A e^{\ii \Omega_A t})F(\bm x-\bm x_A,t)\nonumber\\&\!\!\!\!\!\!\!\!\!\times  \!F(\bm x'-\bm x_B,t') \text{Re}\left(\alpha^*_B\beta_B e^{\ii \Omega_B t'} {\left\langle\cnm{\phi(\bm x,t)}{\phi(\bm x',t')}\right\rangle}\right),
\end{align}
and $\text{d}v=a(t)^3\,\text{d}^3\bm x\, \text{d}t$  is the FRW volume element. 
This expression  is an extension for smeared detectors of the corresponding expression derived in \cite{Comm1,Jonsson:2014lja}. The authors of \cite{Comm1,Jonsson:2014lja} derived it by computing the second order perturbative correction to the transition probability of Bob's detector and then isolating the $\lambda_A\lambda_B$ contributions, which are the leading order contributions to that probability that depend on the presence of Alice.
Remarkably, combining the contributions $\mathcal{O}(\lambda_A\lambda_B)$ from both the $U^{(1)}\rho_0{U^{(1)}}^\dagger$ and  the $U^{(2)}\rho_0$  terms in  \eqref{eq:orders},  the expression for $S$, at leading order, depends on the field only through the expectation value of the field commutator. Since this commutator is a c-number, the result is  independent of the quantum state of the field, which is therefore irrelevant for the results of this section.

We are not going to focus only on communication scenarios where the light signals emitted by Alice     reach Bob, but also on cases when Alice and Bob remain timelike connected,  which constitute the novel communication modality first reported in \cite{Jonsson:2014lja}. In particular, when Alice and Bob are not  lightlike connected,   Alice encodes her message in the quantum fluctuations of the vacuum by switching on her detector $A$  at  $T_{iA}$  and turning it off at a later time $T_{fA}$.    Bob receives Alice's message by probing the quantum fluctuations of the field. In order to do that, he will switch on his detector $B$
at a time  $T_{iB}>T_{fA}$ and turn it off at $T_{fB}>T_{iB}$.

To this end we will compute a lower bound to the capacity of a communication channel between Alice and Bob. We define a simple communication protocol: Alice encodes ``1'' by coupling her detector $A$ to the field, and ``0'' by not coupling it. Later, Bob switches on his detector $B$ and measures its state. If $B$ is excited, Bob interprets a ``1'', and a ``0'' otherwise.  As discussed in \cite{Jonsson:2014lja}, this communication channel constitutes a binary asymmetric channel between Alice and Bob. These channels have the following Shannon capacity \cite{silverman}
\begin{align}\label{capacity0}
C= \frac{-q\, h(p) + p\, h(q)}{q-p} + \log_2 \left( 1+2^{\frac{h(p)-h(q)}{q-p}}\right),
\end{align}
where $h(x)=-x \log_2(x)-(1-x) \log_2(1-x)$, and $p$ and $q$  are  the conditional probabilities of Bob registering a ``1'' if Alice encoded either a ``1'' or a ``0'', respectively.  The difference between $p$ and $q$ is precisely the signaling term $S=p-q$.

 The capacity of this binary asymmetric channel (i.e., the number of bits per use of the channel that Alice transmits to Bob with this protocol) was proven to be non-zero \cite{Jonsson:2014lja}, regardless of the level of noise (within perturbation theory). The leading order contribution to this capacity  is given by
\begin{align}\label{capacity}
C\simeq\lambda^2_A\lambda^2_B\frac{2}{\ln 2}\left(\frac{S_2}{4|\alpha_B||\beta_B|}\right)^2 +\mathcal O(\lambda_\nu^6).
\end{align}

In order to compute both $S$ and $C$, let us first study the form of the field commutator in Eq. \eqref{signaling} for the cosmological spacetime \eqref{eq:FRW}, both for the conformal and minimal couplings of the massless scalar field to the geometry.

\subsection{Field commutator}

We can obtain the commutator from the advanced and retarded Green functions, $G_-$ and $G_+$ respectively,
\begin{align}\label{com_min}
\left\langle\cnm{\phi(x)}{\phi(x')}\right\rangle=\ii \frac{G_-(x,x')-G_+(x,x')}{4\pi},
\end{align}
with  $x=(\bm x,\eta)$.
Here, $G_\pm(x,x')$ are solutions of the wave equation with a point-like source
\begin{align}
(\Box-\xi \mathcal R) G_\pm(x,x')=-\frac{4\pi}{a(\eta)^4} \delta(\eta-\eta')\delta^3(\bm x-\bm x'),
\end{align}
where $\mathcal{R}$ and $\Box$ have been defined in Eqs. \eqref{R} and \eqref{box}.
To compute $G_\pm(x,x')$ it is useful to rescale them,
\begin{align}
G_\pm(x,x')=\frac{g_\pm(x,x')}{a(\eta)a(\eta')},
\end{align}
and  to introduce the function $\hat g$ via  Fourier transform: 
\begin{align}\label{fou}
g_\pm(x,x')&=\pm\frac{\theta(\pm\eta\mp\eta')}{(2\pi)^3}\int \d \bm k\; e^{\ii \bm k (\bm x-\bm x')}\;\hat g(\eta, \eta', k)\nonumber\\
&=\pm\frac{\theta(\pm\eta\mp\eta')}{2\pi^2 R}\int_0^\infty \d k\;k\; \sin(kR)\hat g(\eta, \eta', k),
\end{align}
where $R = \|\bm x-\bm x'\|$. Then $\hat g(\eta, \eta', k)$ is a solution of the ordinary differential equation
\begin{align}\label{diff-pf}
\left(\frac{\d^2}{\d \eta^2}+k^2-(1-6\xi)\frac{\alpha^2-1/4}{\eta^2}\right)\hat g(\eta, \eta', k)=0,
\end{align}
with boundary conditions
\begin{align}\label{bc}
\hat g(\eta= \eta', k)=0, \quad  \frac{\d \hat g}{\d \eta}(\eta= \eta', k)=4\pi,
\end{align}
where $\alpha$ is defined as an auxiliary function of the parameter $w$, \mbox{$\alpha=|(3-3w)/(6w+2)|$}.
Note that in the second line of Eq. \eqref{fou}, we have integrated the angular dependence taking into account that $\hat g$ only depends on $\bm k$ through its modulus.

\subsubsection{Conformal coupling} \label{sec:Conf}

In this case, $\xi=1/6$, and the above differential equation is straightforward to solve. Indeed, this case exhibits conformal invariance and the solution is simply given by a linear combination of plane waves in the conformal time~$\eta$,
\be
\hat g(\eta, \eta', k)=2\pi\ii \left[e^{-\ii k(\eta-\eta')}-e^{\ii k(\eta-\eta')}\right].
\ee
In consequence we get $g_\pm(x,x')=\delta(\eta-\eta'\mp R)/R$, and therefore the commutator reads
\begin{align}\label{com-conf}
& \left\langle\cnm{\phi(\bm x,t)}{\phi(\bm x',t')} \right\rangle =\frac{\ii}{4\pi }\frac{\delta (\Delta\eta+R)-\delta(\Delta\eta-R)}{a(t)a(t')R},
\end{align}
where $\Delta\eta=\eta(t)-\eta(t')$.
This commutator is the same as in Minkowski spacetime, except for overall conformal factors, and vanishes if the events $(\bm x,t)$ and $(\bm x',t')$ are not lightlike connected. Hence, there is no violation of the strong Huygens principle \cite{czapor}: Communication is only possible strictly on the light cone.

\subsubsection{Minimal coupling}

In this case, $\xi=0$, and the differential equation \eqref{diff-pf} has the linearly independent solutions  
\begin{align}
\hat g_{\alpha 1}(\eta,\eta',k)&=\sqrt{|\eta|} J_{\alpha}(k|\eta|),\\
\hat g_{\alpha 2}(\eta,\eta',k)&=\sqrt{|\eta|} Y_{\alpha}(k|\eta|),
\end{align}
where $J_\alpha$ and $Y_\alpha$ denote respectively the Bessel functions of first and second kind  \cite{abramovitz-stegun}.
The solution $\hat g_{\alpha}(\eta, \eta', k)$, where we explicitly denote the dependence on $\alpha$, will be given by a linear combination of $\hat g_{\alpha 1}$ and $\hat g_{\alpha 2}$ with $\eta'$-dependent  coefficients such that the conditions \eqref{bc} are verified. The  result is
\begin{align}\label{sol-diff}
\hat g_{\alpha}(\eta, \eta', k)&=\sqrt{\left|\frac{\eta}{\eta'}\right|} \text{sgn} (\eta')\Big[\mathcal G^{JY}_{\alpha}(\eta,\eta',k)+\mathcal G^{YJ}_{\alpha}(\eta,\eta',k)\Big],
\end{align}
with  sgn being the sign function,  
\begin{align}
\mathcal{G}_{\alpha}^{JY}(\eta,\eta',k) &=\frac{J_{\alpha}(k|\eta|)Y_{\alpha}(k|\eta'|)}{Y_{\alpha}(k|\eta'|)\mathcal L^J_{\alpha}(k|\eta'|)- J_{\alpha}(k|\eta'|)\mathcal L^Y_{\alpha}(k|\eta'|)},\nonumber\\
\mathcal L^{J}_{\alpha}(k|\eta|)&=J_{\alpha-1}(k|\eta|)-J_{\alpha+1}(k|\eta|),
\end{align}
 and $\mathcal{G}^{YJ}_{\alpha}$ and $\mathcal{L}^Y_{\alpha}$ are defined analogously exchanging the Bessel functions  $J_{\alpha}$ and $Y_{\alpha}$.
The  commutator  in \eqref{com_min} is thus given by
\begin{align}\label{com}
\left\langle\cnm{\phi(\bm x,t)}{\phi(\bm x',t')}\right\rangle&=\ii\frac{\theta( -\Delta\eta)-\theta(\Delta\eta)}{\pi^2 a(t)a(t')R}\\
\nonumber&\times\int_0^\infty \d k\sin(kR) \hat g_{\alpha}\big(\eta(t), \eta(t'), k\big).
\end{align}

 In general the above integral has no analytical solution. 
 However, when the combination $\alpha^2-1/4$ in Eq. \eqref{diff-pf} equals 2, namely $\alpha=3/2$, the solutions to that equation are trigonometric functions and the integral can be analytically computed.  This happens for matter dominated ($w=0$) and cosmological constant dominated ($w=-1$) universes.  Explicitly, in those cases we find that
\begin{align}
 J_{3/2}(k|\eta|)&=\sqrt{\frac{2}{\pi k|\eta|}}\left[-\cos(k\eta)+\frac{\sin(k\eta)}{k\eta}\right],\\
 Y_{3/2}(k|\eta|)&= \sqrt{\frac{2}{\pi k|\eta|}}\left[-\sin(k|\eta|)+\frac{\cos(k\eta)}{k|\eta|}\right],\
\end{align}
and then 
\begin{align}
\hat g_{3/2}(\eta, \eta', k)&=\frac{4\pi}{k}\bigg[\left(1+\frac1{k^2|\eta\eta'|}\right)\sin[(k(\eta-\eta')]\nonumber\\&-\frac{\eta-\eta'}{k|\eta\eta'|}\cos[(k(\eta-\eta')]\bigg].
\end{align}
Thus, one obtains
\begin{align}
g_\pm(x,x')
&=\frac{\delta(\eta-\eta'\mp R)}{R}+\frac{\theta(\pm \eta\mp \eta'-R)}{|\eta\eta'|},
\end{align}
and therefore \cite{Poisson:2011nh}
\begin{align}\label{com-min}
\left\langle\cnm{\phi(\bm x,t)}{\phi(\bm x',t')}\right\rangle&=\frac{\ii}{4\pi }\Bigg[\frac{\delta (\Delta\eta+R)-\delta(\Delta\eta-R)}{a(t)a(t')R}\nonumber
 \\*&+\frac{\theta (-\Delta\eta- R) -\theta(\Delta\eta-R) }{a(t)a(t')|\eta(t)\eta(t')|}\Bigg].
\end{align}

In comparison with the commutator of the conformal coupling case \eqref{com-conf}, this commutator  \eqref{com-min}  acquires an extra term  that contains the Heaviside $\theta$-function. As a consequence, the commutator is not confined to the light cone but has support in its \ timelike interior, and therefore gives a non-vanishing contribution to the signaling estimator $S$ even when the events $(\bm x,t)$ and $(\bm x',t')$ are timelike separated. This is the explicit realization of the violation of the strong Huygens principle.
Let us also note that, while the $\delta$-term confined to the light cone decays as the comoving distance $R$ increases, the contribution of the commutator inside the light cone does not decay at all with that separation. It does decay though with the conformal time. Notice that the expression \eqref{com-min} is not covariant since the fields are already evaluated on the worldlines of the detectors.

The above results about the violation of the strong Huygens principle, and about the decay with the comoving distance of the term inducing the violation, have been obtained for the cases $\alpha= 3/2$ (matter dominated and cosmological constant dominated universe). Nevertheless, generically we would arrive to similar conclusions, as the general expression \eqref{com} will be confined on the light cone only in rare situations. This happens for a radiation dominated universe ($w=1/3$), since in this case the last term in Eq.\eqref{diff-pf} vanishes and then we have conformal invariance.  Moreover, even though in general the term violating the strong Huygens principle would decay with the comoving distance $R$ (in this respect the cases $w=0,-1$ are very particular), we expect this decay to be slower than that of the term with support strictly on the light cone. We also note  that in the case of a universe dominated by cosmological constant, a massive scalar field minimally coupled to the geometry and with mass $m=\sqrt {2 |\Lambda|}$ would not violate the strong Huygens principle, as in that case the mass term compensates the curvature term $\xi\mathcal R=2|\Lambda|$ in the wave equation.

 \section{Communication through the Huygens channel in the standard cosmological model}
\label{sec:GR-sig}

For a matter dominated universe  $w=0$, the scale factor and the conformal time as functions of the comoving time $t\in[0,\infty)$ are given by [see Eq. \eqref{class-sol}]
\be  \label{eq:aGR} a(t)= (9\kappa t^2)^{1/3} ,\qquad \eta ( t)= \left(\frac{3t}{ \kappa }\right)^{1/3}.\ee
Here $\kappa=2\pi G \beta /3$  where, recall, $\beta=\rho a^{3}$ is constant. 

In turn, for the cosmological constant case ($w=-1$), $t\in(-\infty,\infty)$ and
\be  \label{eq:aGRcte} a(t)=\tilde\kappa e^{\sqrt{|\Lambda|}t} ,\qquad \eta ( t)=-\frac1{\sqrt{|\Lambda|}\tilde\kappa}e^{-\sqrt{|\Lambda|}t} ,\ee
$\tilde\kappa$ being an integration constant. 

\begin{figure}
\includegraphics[width=0.45\textwidth]{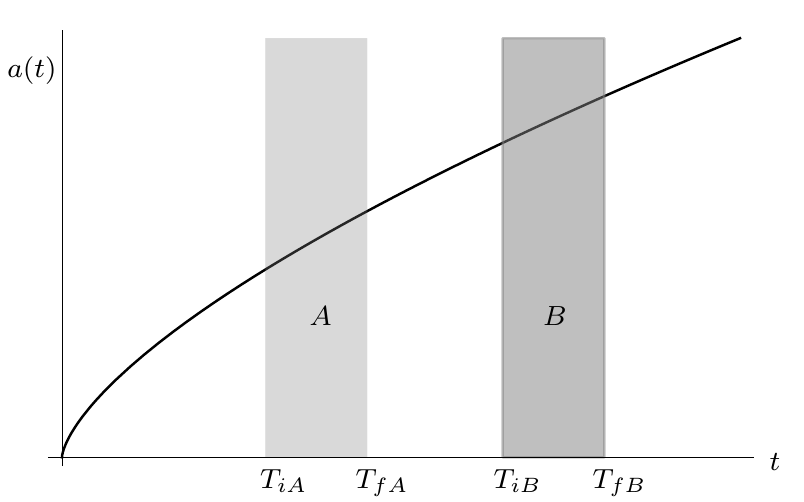}
\caption{Detector's switching strategy for the standard cosmological model, using as example the case $w=0$: Alice's detector interacts with the field in the  interval $[T_{iA},T_{fA}]$, while Bob's detector is turned on at a later time during the  interval $[T_{iB},T_{fB}]$.}
\label{fig:plot_a} 
\end{figure}

In order to compute the signaling estimator $S$,  given in \eqref{S}-\eqref{signaling}, for either a matter dominated universe or a cosmological constant dominated universe, we make use of  either \eqref{eq:aGR}  or \eqref{eq:aGRcte} respectively  to obtain the explicit expression of the commutator, given in Eq. \eqref{com-conf} for the case of the conformal coupling, and in Eq.  \eqref{com-min} for the minimal coupling case. The considered switching strategy for the detectors is illustrated in  Fig.~\ref{fig:plot_a}.

Let us first make the following observation:
Even though the probability of excitation of a sharply switched pointlike detector is UV divergent \cite{Louko:2007mu}, we see that the signaling estimator is UV-safe in the pointlike detector limit ($\sigma\rightarrow0$), even considering sharp switching. Hence, since the pointlike limit is distributionally well behaved, for simplicity we will evaluate the signaling estimator in this limit, taking the abrupt switching function of Eq.~\eqref{eq:switching}. 

In this case, Eq. \eqref{S} reduces to
 \begin{align}\label{sig-point}
S_2\!&=\!4\!\int\! \d t\!\!  \int \!\d t' \chi_A(t)\chi_B(t') \text{Re}(\alpha^*_A\beta_A e^{\ii \Omega_A t})\nonumber\\&\times\text{Re}\left(\alpha^*_B\beta_B e^{\ii \Omega_B t'} { \left\langle\cnm{\phi(\bm x_A,t)}{\phi(\bm x_B,t')}\right\rangle}\right).
\end{align}

Taking into account the explicit form of the commutator \eqref{com-min},  remembering  that the support of the switching function of $A$ precedes the support of the switching function of $B$, and changing the integration variable to conformal time, Eq. \eqref{sig-point} can be recast as 
\begin{equation}
S_2=S_\delta+S_\theta,
\end{equation}
where $S_\delta$ and $S_\theta$ are respectively  the contributions to \eqref{signaling} coming from the Dirac delta and the Heaviside theta terms in \eqref{com-min} (note that  in the case of the conformal coupling we have $S_\theta=0$). They are given by
 \begin{align}\label{Sdelta}
  S_\delta &=
-\frac{1}{\pi R} \int_{-\infty}^{\infty} \d\eta_2 \chi_B(\eta_2)\text{Im}( \alpha^*_B\beta_B e^{\ii \Omega_B t(\eta_2)})\\ &\times\int_{-\infty}^{\infty} \d\eta_1 \chi_A(\eta_1)
\text{Re}(\alpha^*_A\beta_A e^{\ii \Omega_A t(\eta_1)})\delta(\eta_1-\eta_2+ R)\nonumber\\
 S_\theta&=-\frac{1}{\pi}  \int_{-\infty}^{\infty} \d\eta_2 \frac{\chi_B(\eta_2)}{|\eta_2|}
 \text{Im}(\alpha^*_B\beta_B e^{\ii \Omega_B t(\eta_2)})\\ \nonumber
 &\times\int_{-\infty}^{\infty} \d\eta_1 
 \frac{\chi_A(\eta_1)}{|\eta_1|}\text{Re}(\alpha^*_A\beta_A e^{\ii \Omega_A t(\eta_1)})\theta(\eta_2-\eta_1- R).
\end{align}
Expanding the real and imaginary parts of the integrand in terms of trigonometric functions of $\Omega_\nu t$ we  obtain  (with $j=\{\delta,\theta\}$)
\begin{align}\label{S2-j}\nonumber
S_j=\frac{1}{\pi}\Big[&-\text{Re}(\alpha^*_A\beta_A)\text{Re}(\alpha_B\beta^*_B) I_j^{\sin,\cos} \\\nonumber
&+\text{Re}(\alpha^*_A\beta_A)\text{Im}(\alpha_B\beta^*_B) I_j^{\cos,\cos } \\\nonumber
&+ \text{Im}(\alpha^*_A\beta_A)\text{Re}(\alpha_B\beta^*_B)  I_j^{\sin,\sin },\\  
& -\text{Im}(\alpha^*_A\beta_A)\text{Im}(\alpha_B\beta^*_B)  I_j^{\cos,\sin }\Big],  \end{align} 
where we have defined the integrals
\begin{align} 
I_\delta^{\sin,\cos} &= \frac{1}{R}\int_{ \eta_{iB}}^{ \eta_{fB}} \d\eta \chi_A(\eta- R)  \sin[\Omega_B t(\eta) ]  \nonumber\\&\times\cos[\Omega_A t(\eta- R)],\\
I_\theta^{\sin,\cos}&= 
 \int_{ \eta_{iB}}^{ \eta_{fB}} \frac{\d\eta_2}{| \eta_2|} \theta[\text{min}( \eta_{fA},\eta_2-R)- \eta_{iA}]\nonumber \\
 &\times \sin[\Omega_B t(\eta_2)] 
 \int_{ \eta_{iA}}^{\text{min}( \eta_{fA},\eta_2-R)} \frac{\d\eta_1}{|\eta_1 | }  \cos[\Omega_A t(\eta_1) ],
  \end{align}
  and likewise for other combinations of the sine and cosine functions.  Recall that $\eta_{i\nu}=\eta(T_{i\nu})$, $\eta_{f\nu}=\eta(T_{f\nu})$, and $R=\|\vec{x}_A-\vec{x}_B\|$.
For detectors with non-zero gap $\Omega_\nu$, the above integrals do not have closed forms in general, and  we  will  compute them  using numerical methods.

\begin{figure}
\includegraphics[width=0.412\textwidth]{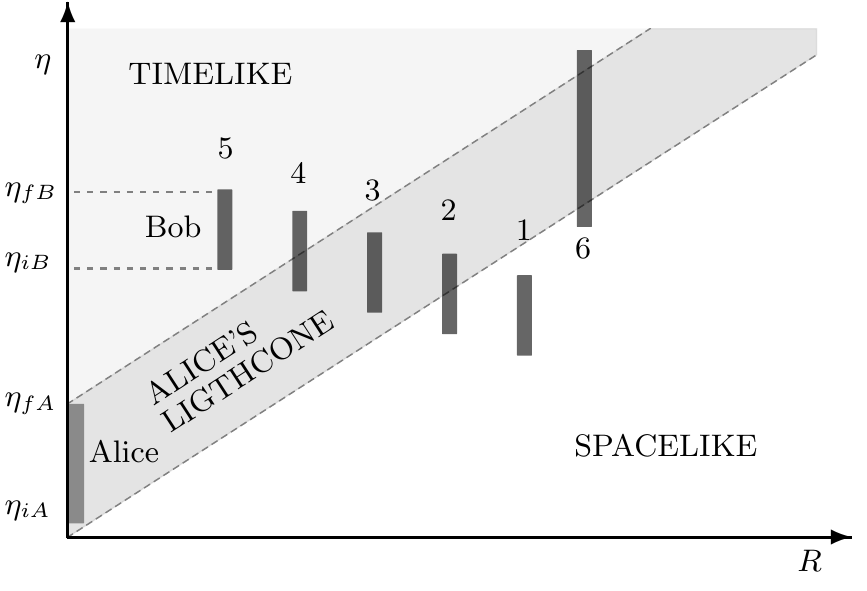}
\caption{Different causal relationships between Alice and Bob's detectors switching periods.   These cases are explicitly specified in Table \ref{tab:table4}.}
\label{fig:cases}
\end{figure}

For the case of zero-gap detectors, $\Omega_\nu=0$, Eq. \eqref{sig-point} admits a fully closed expression. 
The  use  of gapless detectors can be thought as modeling relevant atomic transitions between degenerate (or quasi-degenerate) atomic energy levels, for example, atomic electron spin-flip transitions. Hence, such particle detectors do exist in nature. This kind of transitions happen to actually be very well-modeled by the Unruh-DeWitt model \cite{DeWitt}. 
In this case,  the only non vanishing integrals in \eqref{S2-j} are $I_\delta^{\cos,\cos }$  and $I_\theta^{\cos,\cos }$,  and actually the cosine functions in them trivialize. Thus, we will call them $I_\delta$ and  $I_\theta$ respectively. The   expression for the signaling estimator becomes
\begin{align}\label{S2-point}
S_{2}&=\frac{1}{\pi}  \text{Re}(\alpha^*_A\beta_A)\text{Im}(\alpha_B\beta^*_B) (I_\delta +I_\theta).
\end{align}
In the same fashion as the gapped case, when we have conformal coupling there is no contribution to the  $I_\theta$ term. The explicit  closed  form of $I_\delta$, and of $I_\theta$ for the minimal coupling, depends on the causal relations between Alice's and Bob's detectors.  Recall that we are considering that Alice probes the field before Bob,  then the possible configurations are those depicted in Fig. \ref{fig:cases} and in Table \ref{tab:table4}.

As anticipated in \cite{Blasco:2015eya}, one can then check that the results, for  the different configurations from 1 to 6 in Table \ref{tab:table4}, are respectively
\begin{align}\label{ss}
I_\delta
&=(z_1-z_2)\theta\big(z_1-z_2\big),\\ 
I_\theta&=
\begin{cases}
\ln\left(\frac{\eta_{fA}}
{\eta_{iA}}\right)\ln\left(\frac{\eta_{fB}}
{\eta_{iB}}\right), & \mbox{case $5$},\\
[L(z_1)-L(z_2)+N(z_1)]\theta(z_1-z_2), & \mbox{other cases},
\end{cases} \nonumber
\end{align}
 where  we  have defined
\begin{align}
& L(z)=\ln\left(\dfrac{R(z-1)}{\eta_{iA}}\right)\ln\left(z\right)+\text{Li}_2\left(1-z\right),\\
& N(z)=\ln\left(\dfrac{R(z-1)}{\eta_{iA}}\right)\ln\left(\dfrac{\eta_{fB}}{Rz}\right),\\
& z_1=\frac{\text{min}\left(\eta_{fA}+R,\eta_{fB}\right)}{R},\quad  z_2=\frac{\text{max}\left(\eta_{iA}+R,\eta_{iB}\right)}{R}.
\end{align} 
In the above expressions we have introduced the polylogarithm function $\text{Li}_2(x)$  \cite{abramovitz-stegun}.

 \begin{table} 
\caption{\label{tab:table4}  Cases of causal relationships  between the detectors $A$ and $B$}. 
\begin{ruledtabular}
\begin{tabular}{cl}
\qquad Case&\qquad\qquad Conditions\\
\hline
  \qquad1&$\eta_{fB}\leq \eta_{iA}+R $\\[0.1 cm]
 \qquad 2&$\eta_{iB}<\eta_{iA}+R <\eta_{fB}\leq\eta_{fA}+R \qquad$\\[0.1 cm]
  \qquad3&$\eta_{iB}\geq \eta_{iA}+R$,$\ \eta_{fB}\leq   \eta_{fA}+R $\\[0.1 cm]
\qquad 4 &$\eta_{fB}> \eta_{fA}+R >\eta_{iB}\geq  \eta_{iA}+R $\\[0.1 cm]
\qquad5& $\eta_{iB}\geq \eta_{fA}+R $ \\[0.1 cm]
 \qquad  6&$\eta_{iB}<\eta_{iA}+R $,$\ \eta_{fB}>\eta_{fA}+R$
  \\
\end{tabular}
\end{ruledtabular}
\end{table}

\subsection{Results}

We are now ready to present and discuss the results for the  channel capacity \eqref{capacity}, both for conformal and minimal couplings of the field to the background geometry. 
So far our analysis applies to both  matter-dominated and cosmological-constant dominated universes. Regarding the channel capacity, the difference between these two cases resides in the different form of the function $t(\eta)$ [see \eqref{eq:aGR} and \eqref{eq:aGRcte}].   Although in both cases this function is non-trivial, the results will be qualitatively the same, and we will explicitly compute only one single case.  Taking into account that in the next section we want to carry out the same analysis but considering the effective dynamics of LQC, we will particularize the discussion to the matter dominated universe, which is the one that suffers from a Big Bang singularity in the standard relativistic dynamics, and develops a bounce in the LQC effective dynamics, hence allowing us to compare both scenarios.

For all the plots that we are going to show in this paper we will take the energy density $\rho$ as the scale that sets our unit system. In particular, and for convenience, we take units such that $9\kappa=1$ [see \eqref{eq:aGR}].
 
The signaling estimator \eqref{signaling} helps us to assess the ability of Alice to signal Bob, and the channel capacity \eqref{capacity} provides an estimation of the capacity in bits of the communication channel established between them, being non-zero whenever signaling is possible.
 For the sake of simplicity, we consider that both detectors are switched on for the same amount of time, $\Delta=T_{fA}-T_{iA}= T_{fB}-T_{iB}$. We have selected   their initial  state to be the one that maximizes the channel capacity for the gapless case  (even though we will not restrict to this case), i.e.
\begin{align} \label{eq:conditions}\nonumber
&|\alpha_A|=|\beta_A|=1/\sqrt{2},\\ \nonumber
&\arg(\alpha_A)-\arg(\beta_A)=\pi,\\ 
 &\arg(\alpha_B)-\arg(\beta_B)=\pi/2.
 \end{align}

\begin{figure} 
\includegraphics[width=0.45\textwidth]{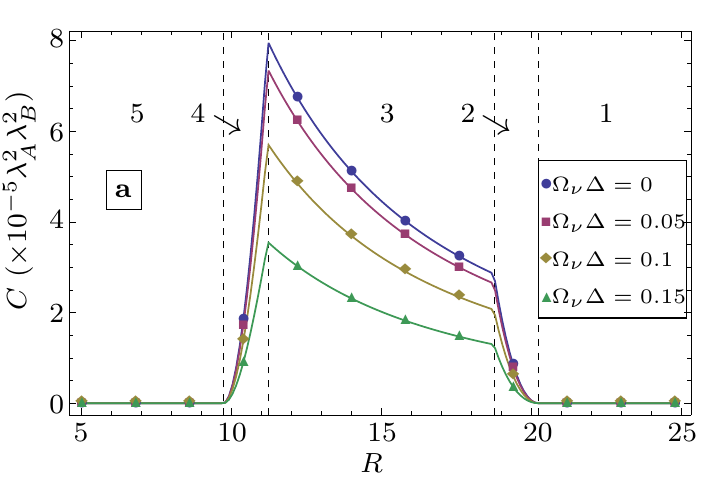} 
\includegraphics[width=0.45\textwidth]{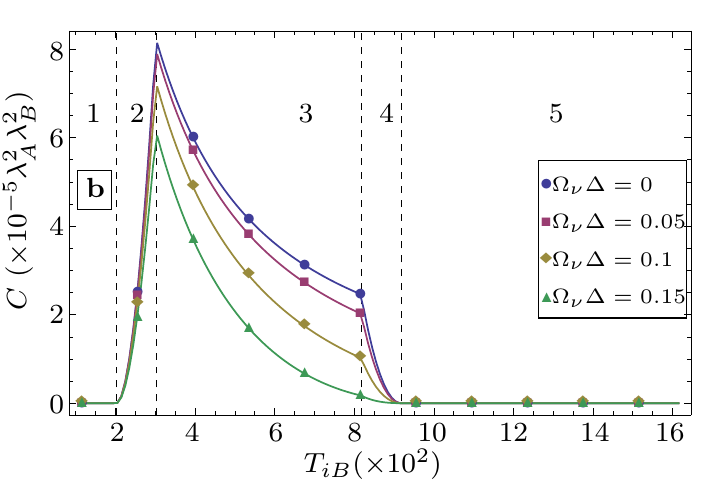}
\caption{Conformal coupling case: a) Variation of the channel capacity with the spatial separation $R$ between detectors. Here, $\Delta=100$, $T_{iA}=5$, and  $T_{iB}=T_{iA}+5\Delta$. b) Variation of the channel capacity with the instant $T_{iB}$ when detector $B$ is switched on. Here, $R=15$, $\Delta=100$, and $T_{iA}=5$. In these plots $R$ and $T_{iB}$ are displayed in the units given by the natural unit scale defined  by $9\kappa=1$.}
\label{fig:Ch_tib_conf_GR} 
\end{figure}

For the case of conformal coupling, Figs. \ref{fig:Ch_tib_conf_GR}-a and \mbox{\ref{fig:Ch_tib_conf_GR}-b} show the variation of the  channel capacity  with the spatial and temporal distance between detectors, respectively, for different values of the detector's energy gap. We have set $\Omega_A=\Omega_B$. The five different regions in these plots correspond to the different causal relationships specified in Fig. \ref{fig:cases}   and in Table \ref{tab:table4}. 
As indicated in Sec. \ref{sec:Conf}, signaling is only allowed for lightlike  events, and therefore the channel capacity vanishes when there is no lightlike connection between the switching periods of $A$ and $B$; namely, either when they are  spacelike separated, which happens when the event $(\vec{x}_B,T_{fB})$ is outside the future light-cone of $(\vec{x}_A,T_{iA})$ (region 1), or when the switching periods are timelike separated, which happens when $(\vec{x}_B,T_{iB})$ is inside the future light-cone of $(\vec{x}_A,T_{fA})$ (region 5). 

Let us focus for a moment in the variation with the spatial separation $R=\|\vec{x}_A-\vec{x}_B\|$, displayed in Fig.~\mbox{\ref{fig:Ch_tib_conf_GR}-a}. In region 5 only timelike connection happens. When events of $A$ and $B$ start being lightlike connected ---the smallest $R$ for which this happens is such that $(\vec{x}_A,T_{fA})$ is lightlike connected  with $(\vec{x}_B,T_{iB})$--- the channel capacity increases (region 4), reaching a maximum when $(\vec{x}_A,T_{fA})$ and $(\vec{x}_B,T_{fB})$ become lightlike connected, because for that configuration all the events of $B$ (while it is switched on) are lightlike connected with events of the switching period of $A$. The decreasing of the channel capacity in region 3 is in part a consequence of the $1/R$ factor in \eqref{Sdelta}. The last point in region 3 corresponds to $(\vec{x}_A,T_{iA})$ and $(\vec{x}_B,T_{iB})$ being lightlike connected. From there onwards many events of the switching period of $B$ are no longer lightlike  connected   with any event of the switching period of $A$ and the channel capacity drastically decreases, until $R$ is so large that the periods when $A$ and $B$ are switched on are strictly spacelike separated. 

An analogous analysis applies for Fig. \ref{fig:Ch_tib_conf_GR}-b, where now $R$ is fixed and the causal relations between $A$ and $B$ depend on their temporal separation, controlled by $B$'s switching instant $T_{iB}$.

\begin{figure}
\includegraphics[width=0.45\textwidth]{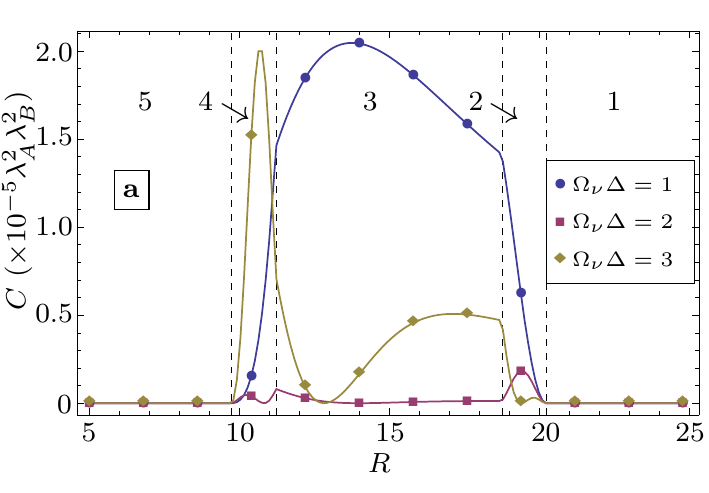} 
\includegraphics[width=0.45\textwidth]{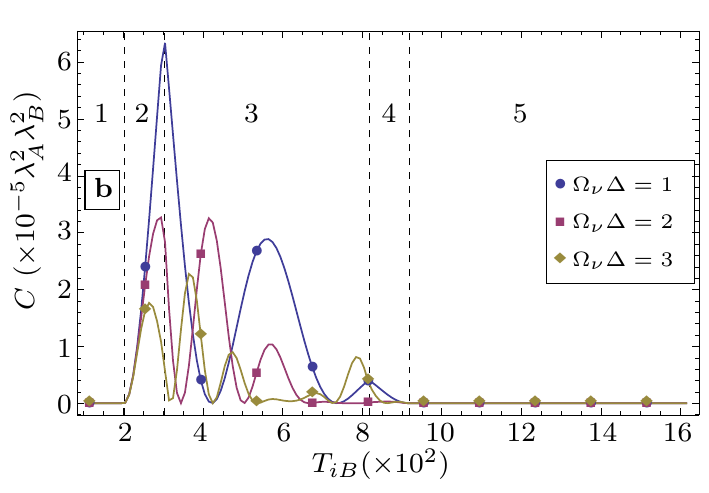}
\caption{Conformal coupling case: a) Variation of the channel capacity with the spatial separation $R$ between detectors. Here, $\Delta=100$, $T_{iA}=5$, and  $T_{iB}=T_{iA}+5\Delta$. b) Variation of the channel capacity with the instant $T_{iB}$ when detector $B$ is switched on. Here, $R=15$, $\Delta=100$, and $T_{iA}=5$. In these plots $R$ and $T_{iB}$ are displayed in the units given by the natural unit scale defined  by $9\kappa=1$.}
\label{fig:Ch_tib_min_GR-omega} 
\end{figure}

Looking now at how the channel capacity behaves as we change the gap of the detector, we see that it is maximum for the gapless detector, and it decreases as we open the gap. This effect is simply due to our choice of initial state for the detector, given in \eqref{eq:conditions}, that precisely maximizes the channel capacity for the gapless case. Changing the initial state would allow us to increase the channel capacity   
up to a certain value by increasing the energy gap. For larger values of $\Omega_\nu$, we see in Fig. \ref{fig:Ch_tib_min_GR-omega} an oscillatory behavior of the channel capacity. Namely, the magnitude of the energy gap modulates  the capacity of the channel.

\begin{figure} 
\includegraphics[width=0.45\textwidth]{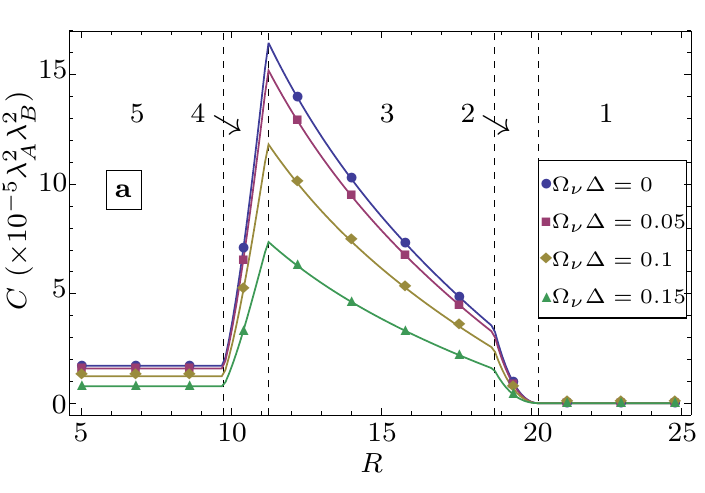} 
\includegraphics[width=0.45\textwidth]{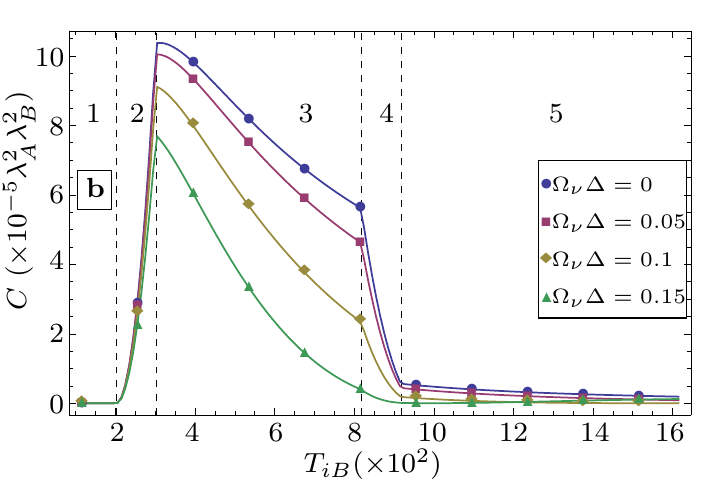}
\caption{Minimal coupling case: a) Variation of the channel capacity with the spatial separation $R$ between detectors. Here, $\Delta=100$, $T_{iA}=5$, and  $T_{iB}=T_{iA}+5\Delta$. b) Variation of the channel capacity with the instant $T_{iB}$ when detector $B$ is switched on. Here, $R=15$, $\Delta=100$, and $T_{iA}=5$. In these plots $R$ and $T_{iB}$ are displayed in the units given by the natural unit scale defined  by $9\kappa=1$.}
\label{fig:Ch_tib_min_GR} 
\end{figure}

Let us now turn attention to the minimal coupling case, for which the behavior of the channel capacity is displayed in Figs. \ref{fig:Ch_tib_min_GR}-a and \ref{fig:Ch_tib_min_GR}-b. We explicitly see the violation of the strong Huygens principle in region 5, that occurs owing to the presence of the $\theta$ term in \eqref{com-min}. This violation opens the door to a non vanishing channel capacity also for timelike separated events. Indeed, in region 5, that corresponds to configurations for which the switching periods of $A$ and $B$ are exclusively timelike connected, the channel capacity does not vanish, in contrast to the conformal coupling case. In other words, even if  Bob's detector is not switched on when the lightlike message reaches him, by switching on his detector at latter times he will still be able to  access information  that is kept encoded in the field.

 In the other regions, from 1 to 4, the capacity is slightly different to that in the conformal coupling case, because now the contribution coming from the $\theta$-term in \eqref{com-min} is non trivially added to that of the $\delta$-term (note that the channel capacity is proportional to the square of the sum of both contributions), which was already present in the conformal coupling case \eqref{com-conf}.
 This $\delta$-term decays with the distance between $A$ and $B$. Therefore, the information transmitted by `rays of light' becomes negligible for long spatial distances. In contrast, the $\theta$-term of \eqref{com-min} does not explicitly decay with $R$, as we can see in region~5 of Fig. \ref{fig:Ch_tib_min_GR}-a.  It decreases though with the time separation between the Big Bang and the switching of both $A$ and $B$. Explicitly, the signaling in region 5 decays logarithmically, as dictated by \eqref{ss}. Remarkably, this decay is slower than the increase of the volume of  Alice's light-cone, and therefore it can in principle be compensated by deploying a big enough number of  separated $B$  receivers in the interior of  Alice's future light-cone in a given time slice. Notice that there could be some entanglement harvesting between these spacelike separated $B$ receivers \cite{reznik,Steeg} correlating their outcomes. Nevertheless, these harvesting correlations can be made small (e.g. turning down $\lambda_B$ while keeping $\lambda_A\lambda_B$ constant) so that the $B$'s become independent users of the channel in good approximation.

\section{The Huygens channel in LQC}
\label{sec:LQC-sig}

Once the channel capacity in the standard cosmological model is obtained,  it is relatively straightforward to obtain the communication capacity of the same protocol between Alice and Bob in the  effective background metric derived from LQC.
This is particularly interesting since LQC predicts a cosmological bounce instead of the initial singularity. This  allows  for the following interesting scenario: What if Alice couples her detector to the field in a pre-bounce time and Bob switches on his detector in a post-bounce era, both far away from the bounce time? 

\begin{figure} 
 \includegraphics[width=0.45\textwidth]{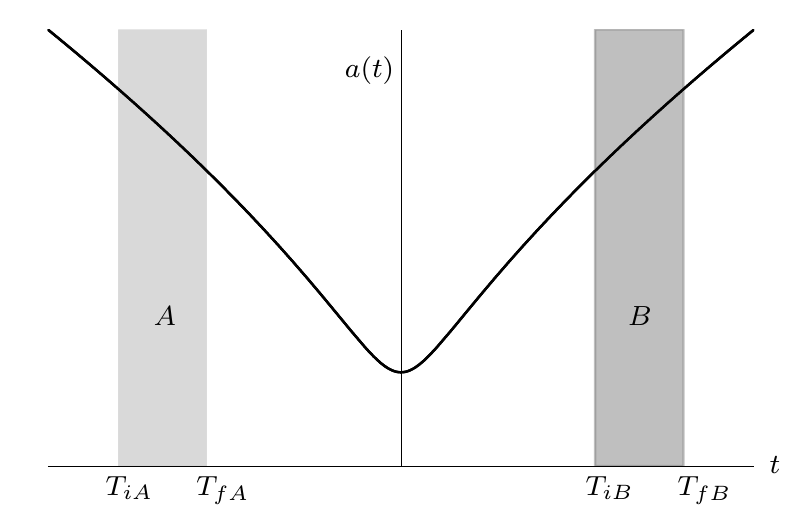}
\caption{Detectors' switching strategy in the LQC scenario: Alice's detector interacts with the field in the contracting universe prior to the bounce in the interval $[T_{iA},T_{fA}]$, with $T_{iA}<T_{fA}<0$, while Bob's detector is turned on during the interval $[T_{iB},T_{fB}]$ at a later time in the expanding  universe that arises after the  bounce, with   $0<T_{iB}<T_{fB}$.}
\label{fig:plot_b} 
\end{figure}

One can think of this scenario with the help of the following cartoon: Imagine an ancient civilization who lived in a pre-bounce era. This civilization mastered physics and therefore they know that their fate is to disappear when, due to the proximity of the bounce, the energy density of the Universe grows higher than the atomic energy bound scales. This civilization does not come to terms with their own disappearance  and thus wants to save their legacy encoding as much information as possible in the quantum field (that will cross through the bounce). They want to do so acting on the field by means of locally coupling particle detectors before the time when those detectors will no longer hold their atomic coherence. One can wonder whether it is possible, at least in principle, to quantify how much information can possibly survive the bounce, and more importantly, how much information could another intelligent species (living in a post-bounce era) recover from the quantum field using particle detectors, once the Universe cools down enough as to allow  again  for the existence of  atoms. The detectors' switching strategy of this scenario is illustrated in Fig. \ref{fig:plot_b}. 

As in the standard cosmological scenario, we will  focus our attention on a matter dominated universe, given by fixing $w=0$ in Eqs. \eqref{eq:scale_LQC}-\eqref{eq:eta_LQC}.
If both Alice and Bob are sufficiently far away from the bounce (each of them at each side of the bounce) the expression of the conformal time in terms of the comoving time, can be approximated as
\begin{align}\label{eta-approx}
 \eta (t)=\left(\frac{3}{\kappa}\right)^\frac13 \text{sgn}(t) \left[\left| t\right|^\frac13+\frac{\sqrt{\pi }\, \Gamma \left(-\frac{1}{6}\right)}{6 ( 6\pi G \rho_\star)^\frac13 \Gamma \left(\frac{1}{3}\right)}\right],
\end{align}
where $\Gamma$ denotes the Gamma function \cite{abramovitz-stegun}.  
Contrary to the  general relativistic  case, now the sign of $\eta$   can be negative. This is well taken into account by means of the absolute values of $\eta(t)$ and $\eta(t')$ that were included in the equations of  Secs. \ref{sec:signaling} and \ref{sec:GR-sig}  in order to make them  directly applicable to this case as well.  Note that the approximation \eqref{eta-approx} allows us to straightforwardly obtain the inverse function $t(\eta)$ needed to compute the signaling estimator and in turn the channel capacity.

\subsection{Results}

\begin{figure}
\includegraphics[width=0.45\textwidth]{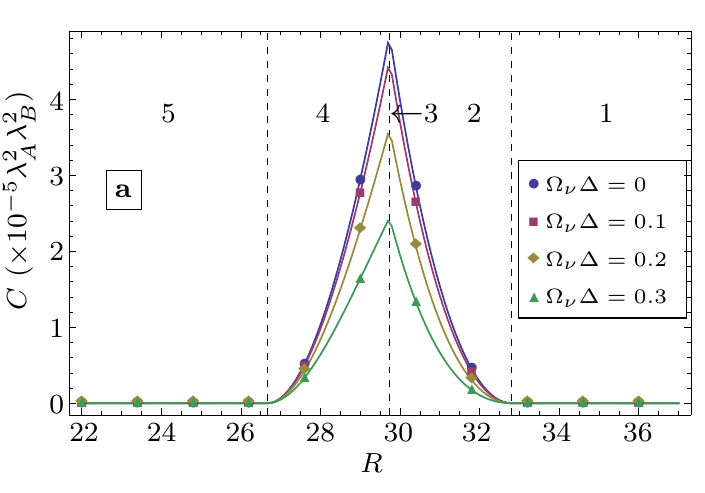}
\includegraphics[width=0.45\textwidth]{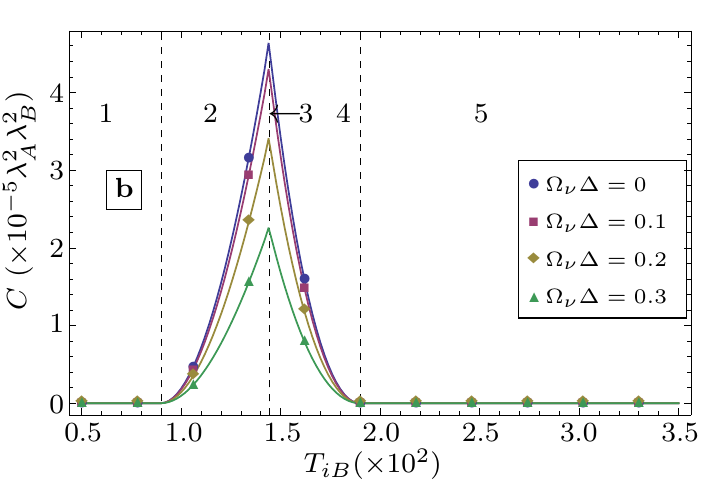}
\caption{Conformal coupling case: a) Variation of the channel capacity with the spatial separation $R$ between detectors. Here, $\Delta=100$ and $-T_{fA}=T_{iB}=175$. b) Variation of the channel capacity with the instant $T_{iB}$ when detector $B$ is switched on. Here, $R=30$, $\Delta=100$,  and $T_{fA}=-T_{iB}$. In these plots $R$ and $T_{iB}$ are displayed in the units given by the natural unit scale defined  by $9\kappa=6\pi G\rho_\star=1$.}
\label{fig:ch_LQC_conf} 
\end{figure}

\begin{figure}
\includegraphics[width=0.45\textwidth]{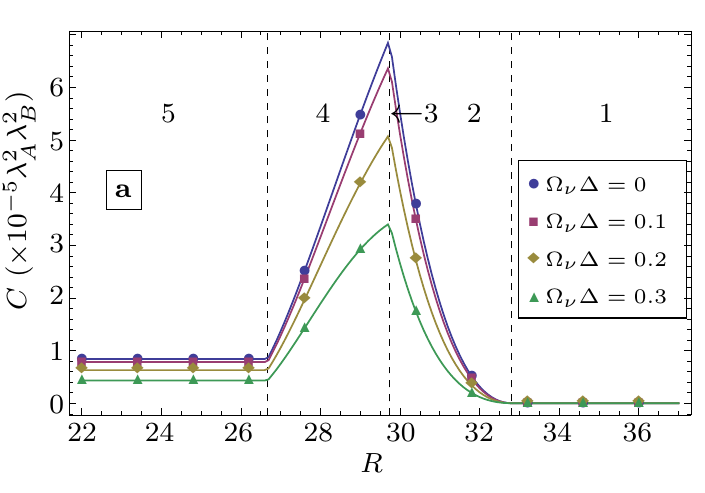}
\includegraphics[width=0.45\textwidth]{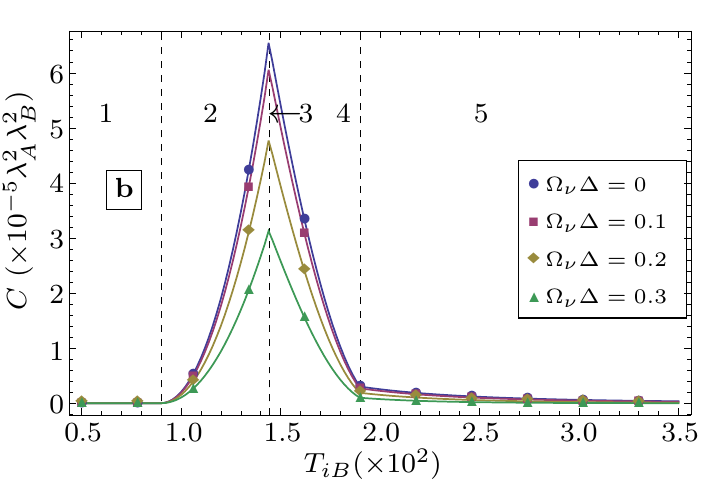}
\caption{Minimal coupling case: a) Variation of the channel capacity with the spatial separation $R$ between detectors. Here, $\Delta=100$, $-T_{fA}=T_{iB}=175$. b) Variation of the channel capacity with the instant $T_{iB}$ when detector $B$ is switched on. Here, $R=30$ and $\Delta=100$,  and $T_{fA}=-T_{iB}$. In these plots $R$ and $T_{iB}$ are displayed in the units given by the natural unit scale defined  by $9\kappa=6\pi G\rho_\star=1$.}
\label{fig:ch_LQC_min} 
\end{figure}

 As initial state of the detectors we still use the one specified in \eqref{eq:conditions} that maximizes the channel capacity. We note that for all the plots regarding the LQC dynamics from now on, we set the LQC scale $\rho_\star=\beta$.
For the  conformal coupling case,
Figs. \ref{fig:ch_LQC_conf}-a  and \ref{fig:ch_LQC_conf}-b show  the variation of the  channel capacity  with the spatial and temporal separation of the detectors for different values of the energy gap. As previously, we have chosen $\Omega_A=\Omega_B$.   The analog plots for the minimal coupling case are displayed in Figs. \ref{fig:ch_LQC_min}-a  and \ref{fig:ch_LQC_min}-b. These plots correspond to a particular setting, in which the switching periods of $A$ and $B$ are symmetric with respect to the bounce, namely we still choose $\Delta=T_{fA}-T_{iA}=T_{fB}-T_{iB}$ but also $T_{iA}=-T_{fB}$, as depicted in Fig. \ref{fig:plot_b} . Notice that this implies $|\eta(T_{iA})|=|\eta(T_{fB})|$ and $|\eta(T_{fA})|=|\eta(T_{iB})|$, and therefore region 3 just collapses into a point where the channel capacity is maximum. Indeed, if $(\vec{x}_A,T_{iA})$ and $(\vec{x}_B,T_{iB})$ are lightlike connected, then $(\vec{x}_A,T_{fA})$ and $(\vec{x}_B,T_{fB})$ are also automatically lightlike connected, and every event of $A$ is lightlike connected  with a event of $B$ and vice versa. On the other hand, in the minimal coupling case, we observe the violation of the strong Huygens principle in an analogous way as in the general relativistic case, and the same conclusions extracted there apply here.

  \section{Mutual information }
 \label{sec:mutual}

In this section we will use the mutual information to  quantify the total amount of correlations (classical and quantum) shared by Alice and Bob. 

The mutual information  of two random variables measures the mutual dependence between them or, being a bit more specific, it measures 
the amount of uncertainty removed from one of the variables after acquiring a single bit of information about the distribution of the other variable. For the quantum states $\rho_{T,A}$ and $\rho_{T,B}$ of the two systems $A$ and $B$, it is defined as
 \be \label{eq:IAB}
  I_{AB}=\mathcal{S}(\rho_{T,A})+\mathcal{S}(\rho_{T,B})-\mathcal{S}(\rho_{T,AB}).\ee
  $\mathcal{S}(\rho)$ denotes the von Neumann entropy 
\be\label{eq:entropyS} \mathcal{S}(\rho) = -\text{tr}(\rho \log_2\rho).\ee
Since the entropy can be interpreted as the missing information about the state, the mutual information can be thought of as a measure of the degree of correlation between  the detectors $A$
and $B$.

Contrary to the channel capacity, the mutual information between $A$ and $B$ after they have interacted with the field does not necessarily vanish when $A$ and $B$ are  spacelike separated. This is a well-known phenomenon known as `vacuum correlation harvesting' (see, e.g. \cite{Pozas-Kerstjens:2015}) that can be traced back to the fact that the field vacuum contains correlations (classical and quantum) between spacelike separated regions, {which are acquired by the detectors through their interaction witht he field}.

We are going to study the mutual information between $A$ and $B$ only in the case of a scalar field conformally coupled to the cosmological background. The reason for this is double: on the one hand it becomes mathematically simpler to focus on a conformally  invariant  case. On the other hand, and more importantly, timelike contributions to the mutual information will be exclusively due to the phenomenon of correlation harvesting, and will not be  affected by  contributions coming from the violation of  the strong Huygens principle, because, as we have already seen, there is no such violation when the coupling is conformal.

In Sec. \ref{rhos-conformal} we already gave the expressions to compute the partial density matrix $\rho_{T,AB}$ of the system formed by the two detectors $A$ and $B$, from which we can in turn compute the partial density matrix of a single detector by tracing out the other detector. In practice, we have computed the integrals \eqref{eq:I1}-\eqref{eq:I3} employing numerical methods, and from them we have obtained $\rho_{T,AB}$ as given in \eqref{eq:rhoAB}. Unlike with the signaling and channel capacity, the mutual information is not well-defined in the limit of point-like detectors due to abrupt-switching related divergences \cite{Louko:2007mu}, and therefore now we cannot take that limit {in a meaningful way}. In order to guarantee that the violation of causality due to the non-vanishing size of the detectors can be neglected, we have chosen the width of the detectors $\sigma$ small enough, as compared to the separation between the detectors. As it can be seen in \cite{Martin-Martinez:2015}, the decay of the causal influence between the two detectors with Gaussian spatial smearing is overexponentially suppressed with the ratio of $\sigma$ and the spatial separation between the center of mass of the detectors.

In the following, we show the results for the mutual information, for the case of the massless field conformally coupled to a matter-dominated universe, both adopting the standard general relativistic dynamics and the effective dynamics derived from LQC.

\begin{figure}
\begin{center}
\includegraphics[width=0.45\textwidth]{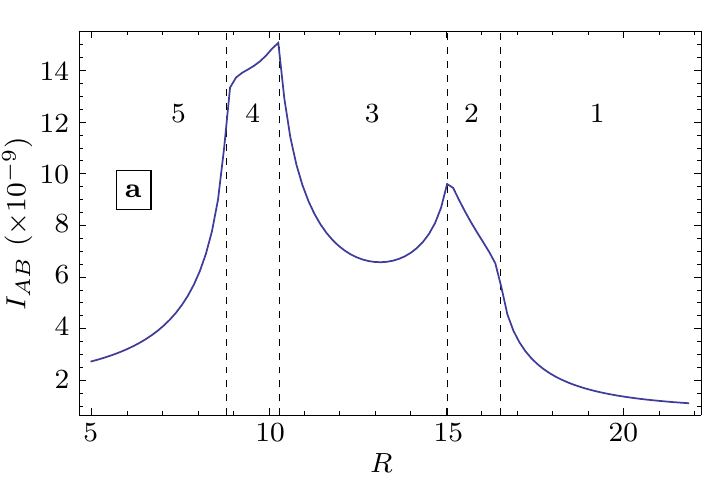}
\includegraphics[width=0.45\textwidth]{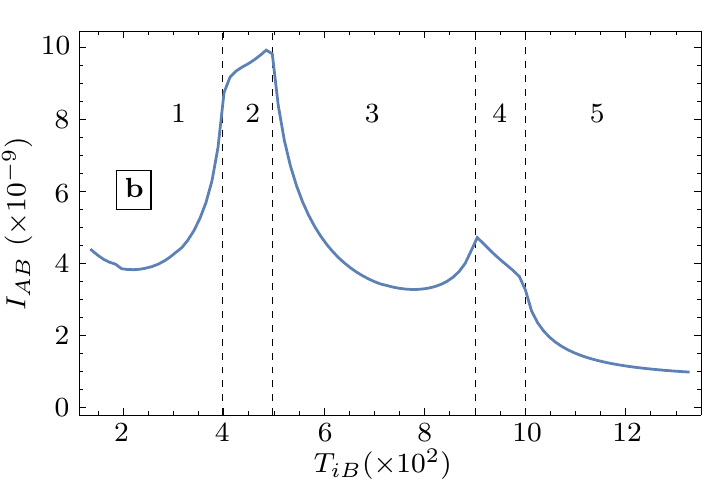}
\caption{Mutual information in the standard cosmological scenario: 
a) Variation with the spatial separation $R$ between detectors. Here, $\lambda_A=\lambda_B=0.01$, $\sigma=0.5$, $\Omega_\nu \Delta=1$, $\Delta=100$, $T_{iA}=25$, and  $T_{iB}=500$. b) Variation with the instant $T_{iB}$ when detector $B$ is switched on. Here, $\lambda_A=\lambda_B=0.01$, $\sigma=0.5$, $R=15$, $\Omega_\nu \Delta=1$, $\Delta=100$, and $T_{iA}=25$. In these plots $R$ and $T_{iB}$ are displayed in the units given by the natural unit scale defined  by $9\kappa=1$.}
\label{fig:IAB_R_GR} 
\end{center}
\end{figure}

For the standard cosmological model,  Figs. \ref{fig:IAB_R_GR}-a and \ref{fig:IAB_R_GR}-b show, respectively, the  behavior of the  mutual information $I_{AB}$ as a function of the detectors' relative spatial  distance $R$, and as a function of the temporal distance, controlled by the switching instant $T_{iB}$ of detector $B$. For simplicity, we have considered that both detectors are switched on during the same amount of time $\Delta$,
and that initially both detectors are in the ground state. The energy gap of the detectors is selected such that $\Omega_\nu \Delta=1$.  Like in previous sections, the different regions from 1 to 5 refer to the corresponding cases in Table \ref{tab:table4} and in Fig. \ref{fig:cases}. 
As shown in Fig. \ref{fig:IAB_R_GR}, while the detectors are only timelike connected (region 5), the mutual information, although non-vanishing, is  small.   Then, it rapidly increases as soon as the  switching period of the  detector $B$ starts to be  lightlike  connected with the  switching period of the  detector $A$ (so they can exchange information), reaching a maximum in region 3, as expected, since this is the optimal configuration in which $B$ is always  lightlike  separated with $A$.   Then, in region 4, this quantity decreases, and as soon as the switching periods of  $B$ and $A$ are no longer causally connected (region 1) it rapidly  tends  to $0$. 
The fact that correlations do not vanish when the detectors are  not lightlike connected, and specially when they are spacelike separated,  stems from the already mentioned  fact that the two detectors `harvest' pre-existing vacuum entanglement and classical correlations \cite{Pozas-Kerstjens:2015}.

\begin{figure}[t]
\includegraphics[width=0.45\textwidth]{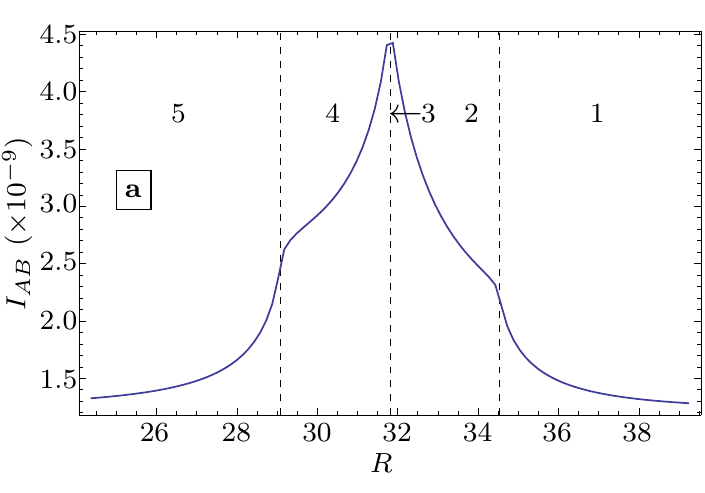}
\includegraphics[width=0.45\textwidth]{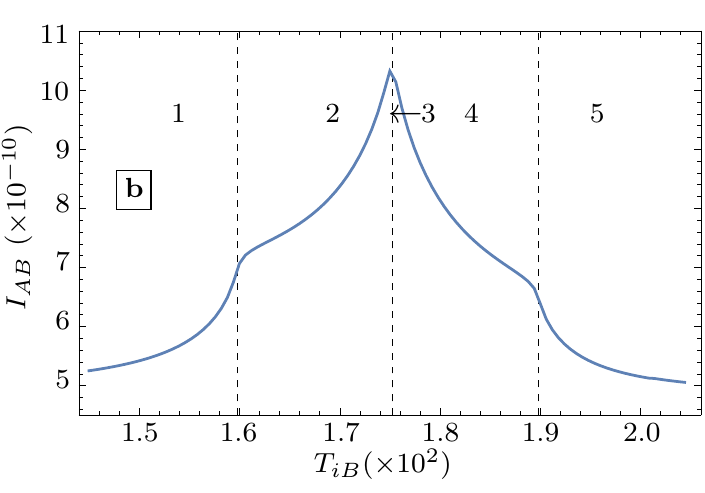}
\caption{Mutual information in the effective LQC scenario: 
a) Variation with the spatial separation $R$ between detectors. Here $\lambda_A=\lambda_B=0.01$, $\sigma=0.5$, $\Omega_\nu \Delta=1$, $\Delta=100$, and $-T_{fA}=T_{iB}=175$. b) Variation with the instant $T_{iB}$ when detector $B$ is switched on. Here,  $\lambda_A=\lambda_B=0.01$, $\sigma=0.5$, $\Omega_\nu \Delta=1$, $R=30$,  $\Delta=30$,  and $T_{fA}=-T_{iB}$. In these plots $R$ and $T_{iB}$ are displayed in the units given by the natural unit scale defined  by $9\kappa=6\pi G \rho_\star=1$.}
\label{fig:IAB_LQC} 
\end{figure}

This is also the case for  the effective LQC model. For this scenario, Fig. \ref{fig:IAB_LQC}-a and Fig. \ref{fig:IAB_LQC}-b depict the variation of the mutual information   as a function of the spatial and temporal separation of the detectors respectively. As in Sec. \ref{sec:LQC-sig}, we have chosen a configuration that is symmetric with respect to the bounce, namely $T_{iA}=-T_{fB}$, with $\Delta=T_{fA}-T_{iA}=T_{fB}-T_{iB}$. 
We  see  correlation  harvesting both in timelike and spatial regions as in the  standard  setting.

\section{Conclusions} \label{sec:con}

We have analyzed the transmission of information from emitters in the early Universe (or in the pre-bounce era in the LQC case) and receivers nowadays. We addressed two relevant questions: 1) In the standard general-relativity scenario, how much information can be transmitted from the early Universe to nowadays? This question was first addressed in \cite{Jonsson:2014lja} and we have broadly generalized here the results on timelike communication via violations of the Strong Huygens principle reported in \cite{Blasco:2015eya}. 2) In the new Loop Quantum Cosmology scenario  we have  investigated  how much information is transmitted through a quantum bounce. To do so we focused on two  quantum information quantities: a lower bound for the channel capacity  between emitter and receiver  and the mutual information between them. Using these estimators we have quantified   two different phenomena: the violation of the strong Huygens principle and the phenomenon of harvesting of classical and quantum correlations from the field \cite{Pozas-Kerstjens:2015}. Additionally, we have characterized the effect of a finite energy gap in the quantum emitter and receiver.

The strong Huygens principle is violated when the propagation is not confined to the lightcone but there is as well a leakage of information towards timelike regions. This is actually the generic thing to happen, as the principle holds only for certain situations such as in Minkowski spacetime or in conformally invariant situations \cite{Ellis,McLenaghan,Sonego:1991sq,czapor}, for which the commutator of a massless field only has support in the lightcone. This  violation of the strong Huygens principle makes the transmission of information possible not only for lightlike connected events but also  for timelike connected ones, the transmission of information being possible even though the receiver cannot receive real quanta from the sender. In these situations, the channel capacity asymptotically decreases for increasing values of conformal time, however  it does not decay with the spatial distance between sender and receiver. This phenomenon was already advanced in  \cite{Blasco:2015eya} for the particular case of detectors with zero energy gap. We have seen that, for each value of the gap, the initial state of the detectors can be adjusted to maximize the capacity of the communication channel. Remarkably, we observe that it is also possible to establish a communication channel in the case the of two detectors  located each on one side of the Big Bounce predicted by  Loop Quantum Cosmology.

We have also studied the mutual information and computed the total amount of correlations (both classical and quantum) shared by emitter and the receiver. 
Comparing our results with the flat spacetime scenario \cite{Pozas-Kerstjens:2015}, we see that the only relevant difference that we observe comes from the fact that the expanding universe changes the shape of the time and distance decay of the ability of the detectors to harvest correlations from the vacuum.

\section{Acknowledgments}
L.J.G. and M.M-B. acknowledge financial support from the  Spanish MICINN/MINECO Project No. FIS2011-30145-C03-02 and its continuation FIS2014-54800-C2-2-P. 
M.M-B. also acknowledges financial support from the Netherlands Organization for Scientific Research (Project No. 62001772). E. M-M. is founded by the NSERC Discovery programme.

\bibliography{biblio}

\begin{thebibliography}{55}%
\makeatletter
\providecommand \@ifxundefined [1]{%
 \@ifx{#1\undefined}
}%
\providecommand \@ifnum [1]{%
 \ifnum #1\expandafter \@firstoftwo
 \else \expandafter \@secondoftwo
 \fi
}%
\providecommand \@ifx [1]{%
 \ifx #1\expandafter \@firstoftwo
 \else \expandafter \@secondoftwo
 \fi
}%
\providecommand \natexlab [1]{#1}%
\providecommand \enquote  [1]{``#1''}%
\providecommand \bibnamefont  [1]{#1}%
\providecommand \bibfnamefont [1]{#1}%
\providecommand \citenamefont [1]{#1}%
\providecommand \href@noop [0]{\@secondoftwo}%
\providecommand \href [0]{\begingroup \@sanitize@url \@href}%
\providecommand \@href[1]{\@@startlink{#1}\@@href}%
\providecommand \@@href[1]{\endgroup#1\@@endlink}%
\providecommand \@sanitize@url [0]{\catcode `\\12\catcode `\$12\catcode
  `\&12\catcode `\#12\catcode `\^12\catcode `\_12\catcode `\%12\relax}%
\providecommand \@@startlink[1]{}%
\providecommand \@@endlink[0]{}%
\providecommand \url  [0]{\begingroup\@sanitize@url \@url }%
\providecommand \@url [1]{\endgroup\@href {#1}{\urlprefix }}%
\providecommand \urlprefix  [0]{URL }%
\providecommand \Eprint [0]{\href }%
\providecommand \doibase [0]{http://dx.doi.org/}%
\providecommand \selectlanguage [0]{\@gobble}%
\providecommand \bibinfo  [0]{\@secondoftwo}%
\providecommand \bibfield  [0]{\@secondoftwo}%
\providecommand \translation [1]{[#1]}%
\providecommand \BibitemOpen [0]{}%
\providecommand \bibitemStop [0]{}%
\providecommand \bibitemNoStop [0]{.\EOS\space}%
\providecommand \EOS [0]{\spacefactor3000\relax}%
\providecommand \BibitemShut  [1]{\csname bibitem#1\endcsname}%
\let\auto@bib@innerbib\@empty
\bibitem [{\citenamefont {Langlois}(2010)}]{Langlois:2010xc}%
  \BibitemOpen
  \bibfield  {author} {\bibinfo {author} {\bibfnamefont {D.}~\bibnamefont
  {Langlois}},\ }\href {\doibase 10.1007/978-3-642-10598-2_1} {\bibfield
  {journal} {\bibinfo  {journal} {Lect. Notes Phys.}\ }\textbf {\bibinfo
  {volume} {800}},\ \bibinfo {pages} {1} (\bibinfo {year} {2010})}\BibitemShut
  {NoStop}%
\bibitem [{\citenamefont {Mukhanov}(2005)}]{Mukhanov:2005sc}%
  \BibitemOpen
  \bibfield  {author} {\bibinfo {author} {\bibfnamefont {V.}~\bibnamefont
  {Mukhanov}},\ }\href@noop {} {\emph {\bibinfo {title} {{Physical foundations
  of cosmology}}}}\ (\bibinfo  {publisher} {Cambridge University Press},\
  \bibinfo {year} {2005})\BibitemShut {NoStop}%
\bibitem [{\citenamefont {Liddle}\ and\ \citenamefont
  {Lyth}(2000)}]{liddle2000cosmological}%
  \BibitemOpen
  \bibfield  {author} {\bibinfo {author} {\bibfnamefont {A.}~\bibnamefont
  {Liddle}}\ and\ \bibinfo {author} {\bibfnamefont {D.}~\bibnamefont {Lyth}},\
  }\href {http://books.google.es/books?id=XmWauPZSovMC} {\emph {\bibinfo
  {title} {Cosmological Inflation and Large-Scale Structure}}}\ (\bibinfo
  {publisher} {Cambridge University Press},\ \bibinfo {year}
  {2000})\BibitemShut {NoStop}%
\bibitem [{\citenamefont {Martin}(2005)}]{Martin:2004um}%
  \BibitemOpen
  \bibfield  {author} {\bibinfo {author} {\bibfnamefont {J.}~\bibnamefont
  {Martin}},\ }\href {\doibase 10.1007/11377306_7} {\bibfield  {journal}
  {\bibinfo  {journal} {Lect. Notes Phys.}\ }\textbf {\bibinfo {volume}
  {669}},\ \bibinfo {pages} {199} (\bibinfo {year} {2005})}\BibitemShut
  {NoStop}%
\bibitem [{\citenamefont {Lahav}\ and\ \citenamefont
  {Liddle}(2014)}]{Lahav:2014vza}%
  \BibitemOpen
  \bibfield  {author} {\bibinfo {author} {\bibfnamefont {O.}~\bibnamefont
  {Lahav}}\ and\ \bibinfo {author} {\bibfnamefont {A.~R.}\ \bibnamefont
  {Liddle}},\ }\href@noop {} {} (\bibinfo {year} {2014}),\ \Eprint
  {http://arxiv.org/abs/ArXiv:1401.1389} {ArXiv:1401.1389} \BibitemShut
  {NoStop}%
\bibitem [{\citenamefont {Ade}\ \emph {et~al.}(2015)\citenamefont {Ade} \emph
  {et~al.}}]{Ade:2015xua}%
  \BibitemOpen
  \bibfield  {author} {\bibinfo {author} {\bibfnamefont {P.~A.~R.}\
  \bibnamefont {Ade}} \emph {et~al.} (\bibinfo {collaboration} {Planck
  Collaboration}),\ }\href@noop {} {} (\bibinfo {year} {2015}),\ \Eprint
  {http://arxiv.org/abs/ArXiv:1502.01589} {ArXiv:1502.01589} \BibitemShut
  {NoStop}%
\bibitem [{\citenamefont {Unruh}(1976)}]{Unruh1976}%
  \BibitemOpen
  \bibfield  {author} {\bibinfo {author} {\bibfnamefont {W.~G.}\ \bibnamefont
  {Unruh}},\ }\href {\doibase 10.1103/PhysRevD.14.870} {\bibfield  {journal}
  {\bibinfo  {journal} {Phys. Rev. D}\ }\textbf {\bibinfo {volume} {14}},\
  \bibinfo {pages} {870} (\bibinfo {year} {1976})}\BibitemShut {NoStop}%
\bibitem [{\citenamefont {Hawking}(1975)}]{Hawking1975}%
  \BibitemOpen
  \bibfield  {author} {\bibinfo {author} {\bibfnamefont {S.~W.}\ \bibnamefont
  {Hawking}},\ }\href@noop {} {\bibfield  {journal} {\bibinfo  {journal} {Comm.
  Math. Phys.}\ }\textbf {\bibinfo {volume} {43}},\ \bibinfo {pages} {199}
  (\bibinfo {year} {1975})}\BibitemShut {NoStop}%
\bibitem [{\citenamefont {Gibbons}\ and\ \citenamefont
  {Hawking}(1977)}]{GibHawking}%
  \BibitemOpen
  \bibfield  {author} {\bibinfo {author} {\bibfnamefont {G.~W.}\ \bibnamefont
  {Gibbons}}\ and\ \bibinfo {author} {\bibfnamefont {S.~W.}\ \bibnamefont
  {Hawking}},\ }\href@noop {} {\bibfield  {journal} {\bibinfo  {journal} {Phys.
  Rev. D}\ }\textbf {\bibinfo {volume} {15}},\ \bibinfo {pages} {2738}
  (\bibinfo {year} {1977})}\BibitemShut {NoStop}%
\bibitem [{\citenamefont {Mart\'{i}n-Mart\'{i}nez}\ and\ \citenamefont
  {Menicucci}(2012)}]{cosmoq}%
  \BibitemOpen
  \bibfield  {author} {\bibinfo {author} {\bibfnamefont {E.}~\bibnamefont
  {Mart\'{i}n-Mart\'{i}nez}}\ and\ \bibinfo {author} {\bibfnamefont {N.~C.}\
  \bibnamefont {Menicucci}},\ }\href {\doibase 10.1088/0264-9381/29/22/224003}
  {\bibfield  {journal} {\bibinfo  {journal} {Class. Quant. Grav.}\ }\textbf
  {\bibinfo {volume} {29}},\ \bibinfo {pages} {224003} (\bibinfo {year}
  {2012})}\BibitemShut {NoStop}%
\bibitem [{\citenamefont {Mart\'in-Mart\'inez}\ and\ \citenamefont
  {Menicucci}(2014)}]{review2}%
  \BibitemOpen
  \bibfield  {author} {\bibinfo {author} {\bibfnamefont {E.}~\bibnamefont
  {Mart\'in-Mart\'inez}}\ and\ \bibinfo {author} {\bibfnamefont {N.~C.}\
  \bibnamefont {Menicucci}},\ }\href
  {http://stacks.iop.org/0264-9381/31/i=21/a=214001} {\bibfield  {journal}
  {\bibinfo  {journal} {Class. Quant. Grav.}\ }\textbf {\bibinfo {volume}
  {31}},\ \bibinfo {pages} {214001} (\bibinfo {year} {2014})}\BibitemShut
  {NoStop}%
\bibitem [{\citenamefont {{Ver Steeg}}\ and\ \citenamefont
  {Menicucci}(2009)}]{Steeg}%
  \BibitemOpen
  \bibfield  {author} {\bibinfo {author} {\bibfnamefont {G.}~\bibnamefont {{Ver
  Steeg}}}\ and\ \bibinfo {author} {\bibfnamefont {N.~C.}\ \bibnamefont
  {Menicucci}},\ }\href@noop {} {\bibfield  {journal} {\bibinfo  {journal}
  {Phys. Rev. D}\ }\textbf {\bibinfo {volume} {79}},\ \bibinfo {pages} {044027}
  (\bibinfo {year} {2009})}\BibitemShut {NoStop}%
\bibitem [{\citenamefont {Reznik}\ \emph {et~al.}(2005)\citenamefont {Reznik},
  \citenamefont {Retzker},\ and\ \citenamefont {Silman}}]{reznik}%
  \BibitemOpen
  \bibfield  {author} {\bibinfo {author} {\bibfnamefont {B.}~\bibnamefont
  {Reznik}}, \bibinfo {author} {\bibfnamefont {A.}~\bibnamefont {Retzker}}, \
  and\ \bibinfo {author} {\bibfnamefont {J.}~\bibnamefont {Silman}},\
  }\href@noop {} {\bibfield  {journal} {\bibinfo  {journal} {Phys. Rev. A}\
  }\textbf {\bibinfo {volume} {71}},\ \bibinfo {eid} {042104} (\bibinfo {year}
  {2005})}\BibitemShut {NoStop}%
\bibitem [{\citenamefont {Olson}\ and\ \citenamefont
  {Ralph}(2011)}]{Olson2011}%
  \BibitemOpen
  \bibfield  {author} {\bibinfo {author} {\bibfnamefont {S.~J.}\ \bibnamefont
  {Olson}}\ and\ \bibinfo {author} {\bibfnamefont {T.~C.}\ \bibnamefont
  {Ralph}},\ }\href {\doibase 10.1103/PhysRevLett.106.110404} {\bibfield
  {journal} {\bibinfo  {journal} {Phys. Rev. Lett.}\ }\textbf {\bibinfo
  {volume} {106}},\ \bibinfo {pages} {110404} (\bibinfo {year}
  {2011})}\BibitemShut {NoStop}%
\bibitem [{\citenamefont {Mart\'{i}n-Mart\'{i}nez}\ \emph
  {et~al.}(2012)\citenamefont {Mart\'{i}n-Mart\'{i}nez}, \citenamefont
  {Garay},\ and\ \citenamefont {Leon}}]{Collapse2}%
  \BibitemOpen
  \bibfield  {author} {\bibinfo {author} {\bibfnamefont {E.}~\bibnamefont
  {Mart\'{i}n-Mart\'{i}nez}}, \bibinfo {author} {\bibfnamefont {L.~J.}\
  \bibnamefont {Garay}}, \ and\ \bibinfo {author} {\bibfnamefont
  {J.}~\bibnamefont {Leon}},\ }\href@noop {} {\bibfield  {journal} {\bibinfo
  {journal} {Class. Quant. Grav.}\ }\textbf {\bibinfo {volume} {29}},\ \bibinfo
  {pages} {224006} (\bibinfo {year} {2012})}\BibitemShut {NoStop}%
\bibitem [{\citenamefont {Jonsson}\ \emph {et~al.}(2014)\citenamefont
  {Jonsson}, \citenamefont {Mart\'{i}n-Mart\'{i}nez},\ and\ \citenamefont
  {Kempf}}]{Comm1}%
  \BibitemOpen
  \bibfield  {author} {\bibinfo {author} {\bibfnamefont {R.~H.}\ \bibnamefont
  {Jonsson}}, \bibinfo {author} {\bibfnamefont {E.}~\bibnamefont
  {Mart\'{i}n-Mart\'{i}nez}}, \ and\ \bibinfo {author} {\bibfnamefont
  {A.}~\bibnamefont {Kempf}},\ }\href {\doibase 10.1103/PhysRevA.89.022330}
  {\bibfield  {journal} {\bibinfo  {journal} {Phys. Rev. A}\ }\textbf {\bibinfo
  {volume} {89}},\ \bibinfo {pages} {022330} (\bibinfo {year}
  {2014})}\BibitemShut {NoStop}%
\bibitem [{\citenamefont {Jonsson}\ \emph {et~al.}(2015)\citenamefont
  {Jonsson}, \citenamefont {Mart\'{i}n-Mart\'{i}nez},\ and\ \citenamefont
  {Kempf}}]{Jonsson:2014lja}%
  \BibitemOpen
  \bibfield  {author} {\bibinfo {author} {\bibfnamefont {R.~H.}\ \bibnamefont
  {Jonsson}}, \bibinfo {author} {\bibfnamefont {E.}~\bibnamefont
  {Mart\'{i}n-Mart\'{i}nez}}, \ and\ \bibinfo {author} {\bibfnamefont
  {A.}~\bibnamefont {Kempf}},\ }\href {\doibase 10.1103/PhysRevLett.114.110505}
  {\bibfield  {journal} {\bibinfo  {journal} {Phys. Rev. Lett.}\ }\textbf
  {\bibinfo {volume} {114}},\ \bibinfo {pages} {110505} (\bibinfo {year}
  {2015})}\BibitemShut {NoStop}%
\bibitem [{\citenamefont {Ashtekar}\ \emph {et~al.}(2006)\citenamefont
  {Ashtekar}, \citenamefont {Pawlowski},\ and\ \citenamefont
  {Singh}}]{Ashtekar:2006wn}%
  \BibitemOpen
  \bibfield  {author} {\bibinfo {author} {\bibfnamefont {A.}~\bibnamefont
  {Ashtekar}}, \bibinfo {author} {\bibfnamefont {T.}~\bibnamefont {Pawlowski}},
  \ and\ \bibinfo {author} {\bibfnamefont {P.}~\bibnamefont {Singh}},\ }\href
  {\doibase 10.1103/PhysRevD.74.084003} {\bibfield  {journal} {\bibinfo
  {journal} {Phys. Rev. D}\ }\textbf {\bibinfo {volume} {74}},\ \bibinfo
  {pages} {084003} (\bibinfo {year} {2006})}\BibitemShut {NoStop}%
\bibitem [{\citenamefont {Bojowald}(2008)}]{Bojowald:2008zzb}%
  \BibitemOpen
  \bibfield  {author} {\bibinfo {author} {\bibfnamefont {M.}~\bibnamefont
  {Bojowald}},\ }\href@noop {} {\bibfield  {journal} {\bibinfo  {journal}
  {Living Rev. Rel.}\ }\textbf {\bibinfo {volume} {11}},\ \bibinfo {pages} {4}
  (\bibinfo {year} {2008})}\BibitemShut {NoStop}%
\bibitem [{\citenamefont {Banerjee}\ \emph {et~al.}(2012)\citenamefont
  {Banerjee}, \citenamefont {Calcagni},\ and\ \citenamefont
  {Mart\'in-Benito}}]{Banerjee:2011qu}%
  \BibitemOpen
  \bibfield  {author} {\bibinfo {author} {\bibfnamefont {K.}~\bibnamefont
  {Banerjee}}, \bibinfo {author} {\bibfnamefont {G.}~\bibnamefont {Calcagni}},
  \ and\ \bibinfo {author} {\bibfnamefont {M.}~\bibnamefont
  {Mart\'in-Benito}},\ }\href {\doibase 10.3842/SIGMA.2012.016} {\bibfield
  {journal} {\bibinfo  {journal} {SIGMA}\ }\textbf {\bibinfo {volume} {8}},\
  \bibinfo {pages} {016} (\bibinfo {year} {2012})}\BibitemShut {NoStop}%
\bibitem [{\citenamefont {Ashtekar}\ and\ \citenamefont
  {Singh}(2011)}]{Ashtekar:2011ni}%
  \BibitemOpen
  \bibfield  {author} {\bibinfo {author} {\bibfnamefont {A.}~\bibnamefont
  {Ashtekar}}\ and\ \bibinfo {author} {\bibfnamefont {P.}~\bibnamefont
  {Singh}},\ }\href {\doibase 10.1088/0264-9381/28/21/213001} {\bibfield
  {journal} {\bibinfo  {journal} {Class. Quant. Grav.}\ }\textbf {\bibinfo
  {volume} {28}},\ \bibinfo {pages} {213001} (\bibinfo {year}
  {2011})}\BibitemShut {NoStop}%
\bibitem [{\citenamefont {Garay}\ \emph {et~al.}(2014)\citenamefont {Garay},
  \citenamefont {Mart\'in-Benito},\ and\ \citenamefont
  {Mart\'in-Mart\'inez}}]{Garay:2013dya}%
  \BibitemOpen
  \bibfield  {author} {\bibinfo {author} {\bibfnamefont {L.~J.}\ \bibnamefont
  {Garay}}, \bibinfo {author} {\bibfnamefont {M.}~\bibnamefont
  {Mart\'in-Benito}}, \ and\ \bibinfo {author} {\bibfnamefont {E.}~\bibnamefont
  {Mart\'in-Mart\'inez}},\ }\href {\doibase 10.1103/PhysRevD.89.043510}
  {\bibfield  {journal} {\bibinfo  {journal} {Phys. Rev. D}\ }\textbf {\bibinfo
  {volume} {89}},\ \bibinfo {pages} {043510} (\bibinfo {year}
  {2014})}\BibitemShut {NoStop}%
\bibitem [{\citenamefont {Valentini}(1991)}]{Valentini1991321}%
  \BibitemOpen
  \bibfield  {author} {\bibinfo {author} {\bibfnamefont {A.}~\bibnamefont
  {Valentini}},\ }\href {\doibase
  http://dx.doi.org/10.1016/0375-9601(91)90952-5} {\bibfield  {journal}
  {\bibinfo  {journal} {Phys. Lett. A}\ }\textbf {\bibinfo {volume} {153}},\
  \bibinfo {pages} {321 } (\bibinfo {year} {1991})}\BibitemShut {NoStop}%
\bibitem [{\citenamefont {Salton}\ \emph {et~al.}(2015)\citenamefont {Salton},
  \citenamefont {Mann},\ and\ \citenamefont {Menicucci}}]{Nicklast}%
  \BibitemOpen
  \bibfield  {author} {\bibinfo {author} {\bibfnamefont {G.}~\bibnamefont
  {Salton}}, \bibinfo {author} {\bibfnamefont {R.~B.}\ \bibnamefont {Mann}}, \
  and\ \bibinfo {author} {\bibfnamefont {N.~C.}\ \bibnamefont {Menicucci}},\
  }\href {\doibase doi:10.1088/1367-2630/17/3/035001} {\bibfield  {journal}
  {\bibinfo  {journal} {New J. Phys.}\ }\textbf {\bibinfo {volume} {17}},\
  \bibinfo {pages} {035001} (\bibinfo {year} {2015})}\BibitemShut {NoStop}%
\bibitem [{\citenamefont {Mart\'in-Mart\'inez}\ \emph
  {et~al.}(2013)\citenamefont {Mart\'in-Mart\'inez}, \citenamefont {Brown},
  \citenamefont {Donnelly},\ and\ \citenamefont {Kempf}}]{farming}%
  \BibitemOpen
  \bibfield  {author} {\bibinfo {author} {\bibfnamefont {E.}~\bibnamefont
  {Mart\'in-Mart\'inez}}, \bibinfo {author} {\bibfnamefont {E.~G.}\
  \bibnamefont {Brown}}, \bibinfo {author} {\bibfnamefont {W.}~\bibnamefont
  {Donnelly}}, \ and\ \bibinfo {author} {\bibfnamefont {A.}~\bibnamefont
  {Kempf}},\ }\href {\doibase 10.1103/PhysRevA.88.052310} {\bibfield  {journal}
  {\bibinfo  {journal} {Phys. Rev. A}\ }\textbf {\bibinfo {volume} {88}},\
  \bibinfo {pages} {052310} (\bibinfo {year} {2013})}\BibitemShut {NoStop}%
\bibitem [{\citenamefont {Ellis}\ and\ \citenamefont {Sciama}(1972)}]{Ellis}%
  \BibitemOpen
  \bibfield  {author} {\bibinfo {author} {\bibfnamefont {G.~F.~R.}\
  \bibnamefont {Ellis}}\ and\ \bibinfo {author} {\bibfnamefont {D.~W.}\
  \bibnamefont {Sciama}},\ }\href@noop {} {\emph {\bibinfo {title} {in L.
  O'Raifeartaigh, ed., General Relativity, Papers in Honour of J. L. Synge}}}\
  (\bibinfo  {publisher} {Oxford: Clarendon Press},\ \bibinfo {year}
  {1972})\BibitemShut {NoStop}%
\bibitem [{\citenamefont {McLenaghan}(1974)}]{McLenaghan}%
  \BibitemOpen
  \bibfield  {author} {\bibinfo {author} {\bibfnamefont {R.}~\bibnamefont
  {McLenaghan}},\ }\href@noop {} {\bibfield  {journal} {\bibinfo  {journal}
  {Ann. Inst. H. Poincare}\ }\textbf {\bibinfo {volume} {20}},\ \bibinfo
  {pages} {153} (\bibinfo {year} {1974})}\BibitemShut {NoStop}%
\bibitem [{\citenamefont {Sonego}\ and\ \citenamefont
  {Faraoni}(1992)}]{Sonego:1991sq}%
  \BibitemOpen
  \bibfield  {author} {\bibinfo {author} {\bibfnamefont {S.}~\bibnamefont
  {Sonego}}\ and\ \bibinfo {author} {\bibfnamefont {V.}~\bibnamefont
  {Faraoni}},\ }\href {\doibase 10.1063/1.529798} {\bibfield  {journal}
  {\bibinfo  {journal} {J. Math. Phys.}\ }\textbf {\bibinfo {volume} {33}},\
  \bibinfo {pages} {625} (\bibinfo {year} {1992})}\BibitemShut {NoStop}%
\bibitem [{\citenamefont {Czapor}\ and\ \citenamefont
  {McLenaghan}(2008)}]{czapor}%
  \BibitemOpen
  \bibfield  {author} {\bibinfo {author} {\bibfnamefont {S.}~\bibnamefont
  {Czapor}}\ and\ \bibinfo {author} {\bibfnamefont {R.}~\bibnamefont
  {McLenaghan}},\ }\href
  {http://www.actaphys.uj.edu.pl/_old/sup1/pdf/s1p0055.pdf} {\bibfield
  {journal} {\bibinfo  {journal} {Acta. Phys. Pol. B Proc. Suppl.}\ }\textbf
  {\bibinfo {volume} {1}},\ \bibinfo {pages} {55} (\bibinfo {year}
  {2008})}\BibitemShut {NoStop}%
\bibitem [{\citenamefont {Blanchet}\ and\ \citenamefont
  {Damour}(1988)}]{Blanchet:1987wq}%
  \BibitemOpen
  \bibfield  {author} {\bibinfo {author} {\bibfnamefont {L.}~\bibnamefont
  {Blanchet}}\ and\ \bibinfo {author} {\bibfnamefont {T.}~\bibnamefont
  {Damour}},\ }\href {\doibase 10.1103/PhysRevD.37.1410} {\bibfield  {journal}
  {\bibinfo  {journal} {Phys. Rev. D}\ }\textbf {\bibinfo {volume} {37}},\
  \bibinfo {pages} {1410} (\bibinfo {year} {1988})}\BibitemShut {NoStop}%
\bibitem [{\citenamefont {Blanchet}\ and\ \citenamefont
  {Damour}(1992)}]{Blanchet:1992}%
  \BibitemOpen
  \bibfield  {author} {\bibinfo {author} {\bibfnamefont {L.}~\bibnamefont
  {Blanchet}}\ and\ \bibinfo {author} {\bibfnamefont {T.}~\bibnamefont
  {Damour}},\ }\href {\doibase 10.1103/PhysRevD.46.4304} {\bibfield  {journal}
  {\bibinfo  {journal} {Phys. Rev. D}\ }\textbf {\bibinfo {volume} {46}},\
  \bibinfo {pages} {4304} (\bibinfo {year} {1992})}\BibitemShut {NoStop}%
\bibitem [{\citenamefont {Bombelli}\ and\ \citenamefont
  {Sonego}(1994)}]{Bombelli:1994rh}%
  \BibitemOpen
  \bibfield  {author} {\bibinfo {author} {\bibfnamefont {L.}~\bibnamefont
  {Bombelli}}\ and\ \bibinfo {author} {\bibfnamefont {S.}~\bibnamefont
  {Sonego}},\ }\href {\doibase 10.1088/0305-4470/27/21/033} {\bibfield
  {journal} {\bibinfo  {journal} {J. Phys.}\ }\textbf {\bibinfo {volume}
  {A27}},\ \bibinfo {pages} {7177} (\bibinfo {year} {1994})}\BibitemShut
  {NoStop}%
\bibitem [{\citenamefont {Gundlach}\ \emph {et~al.}(1994)\citenamefont
  {Gundlach}, \citenamefont {Price},\ and\ \citenamefont {Pullin}}]{Pullin}%
  \BibitemOpen
  \bibfield  {author} {\bibinfo {author} {\bibfnamefont {C.}~\bibnamefont
  {Gundlach}}, \bibinfo {author} {\bibfnamefont {R.~H.}\ \bibnamefont {Price}},
  \ and\ \bibinfo {author} {\bibfnamefont {J.}~\bibnamefont {Pullin}},\ }\href
  {\doibase 10.1103/PhysRevD.49.890} {\bibfield  {journal} {\bibinfo  {journal}
  {Phys. Rev. D}\ }\textbf {\bibinfo {volume} {49}},\ \bibinfo {pages} {890}
  (\bibinfo {year} {1994})}\BibitemShut {NoStop}%
\bibitem [{\citenamefont {Hod}(2000)}]{Hod:1999ci}%
  \BibitemOpen
  \bibfield  {author} {\bibinfo {author} {\bibfnamefont {S.}~\bibnamefont
  {Hod}},\ }\href {\doibase 10.1103/PhysRevLett.84.10} {\bibfield  {journal}
  {\bibinfo  {journal} {Phys. Rev. Lett.}\ }\textbf {\bibinfo {volume} {84}},\
  \bibinfo {pages} {10} (\bibinfo {year} {2000})}\BibitemShut {NoStop}%
\bibitem [{\citenamefont {Faraoni}\ and\ \citenamefont
  {Sonego}(1992)}]{Faraoni:1991xe}%
  \BibitemOpen
  \bibfield  {author} {\bibinfo {author} {\bibfnamefont {V.}~\bibnamefont
  {Faraoni}}\ and\ \bibinfo {author} {\bibfnamefont {S.}~\bibnamefont
  {Sonego}},\ }\href {\doibase 10.1016/0375-9601(92)90744-7} {\bibfield
  {journal} {\bibinfo  {journal} {Phys. Lett. A}\ }\textbf {\bibinfo {volume}
  {170}},\ \bibinfo {pages} {413} (\bibinfo {year} {1992})}\BibitemShut
  {NoStop}%
\bibitem [{\citenamefont {Faraoni}\ and\ \citenamefont
  {Gunzig}(1999)}]{Faraoni:1999us}%
  \BibitemOpen
  \bibfield  {author} {\bibinfo {author} {\bibfnamefont {V.}~\bibnamefont
  {Faraoni}}\ and\ \bibinfo {author} {\bibfnamefont {E.}~\bibnamefont
  {Gunzig}},\ }\href {\doibase 10.1142/S021827189900016X} {\bibfield  {journal}
  {\bibinfo  {journal} {Int. J. Mod. Phys. D}\ }\textbf {\bibinfo {volume}
  {8}},\ \bibinfo {pages} {177} (\bibinfo {year} {1999})}\BibitemShut {NoStop}%
\bibitem [{\citenamefont {Blasco}\ \emph {et~al.}(2015)\citenamefont {Blasco},
  \citenamefont {Garay}, \citenamefont {Mart\'in-Benito},\ and\ \citenamefont
  {Mart\'in-Mart\'inez}}]{Blasco:2015eya}%
  \BibitemOpen
  \bibfield  {author} {\bibinfo {author} {\bibfnamefont {A.}~\bibnamefont
  {Blasco}}, \bibinfo {author} {\bibfnamefont {L.~J.}\ \bibnamefont {Garay}},
  \bibinfo {author} {\bibfnamefont {M.}~\bibnamefont {Mart\'in-Benito}}, \ and\
  \bibinfo {author} {\bibfnamefont {E.}~\bibnamefont {Mart\'in-Mart\'inez}},\
  }\href {\doibase 10.1103/PhysRevLett.114.141103} {\bibfield  {journal}
  {\bibinfo  {journal} {Phys. Rev. Lett.}\ }\textbf {\bibinfo {volume} {114}},\
  \bibinfo {pages} {141103} (\bibinfo {year} {2015})}\BibitemShut {NoStop}%
\bibitem [{\citenamefont {Taveras}(2008)}]{Taveras:2008ke}%
  \BibitemOpen
  \bibfield  {author} {\bibinfo {author} {\bibfnamefont {V.}~\bibnamefont
  {Taveras}},\ }\href {\doibase 10.1103/PhysRevD.78.064072} {\bibfield
  {journal} {\bibinfo  {journal} {Phys. Rev. D}\ }\textbf {\bibinfo {volume}
  {78}},\ \bibinfo {pages} {064072} (\bibinfo {year} {2008})}\BibitemShut
  {NoStop}%
\bibitem [{\citenamefont {Ashtekar}\ \emph {et~al.}(2008)\citenamefont
  {Ashtekar}, \citenamefont {Corichi},\ and\ \citenamefont
  {Singh}}]{Ashtekar:2007em}%
  \BibitemOpen
  \bibfield  {author} {\bibinfo {author} {\bibfnamefont {A.}~\bibnamefont
  {Ashtekar}}, \bibinfo {author} {\bibfnamefont {A.}~\bibnamefont {Corichi}}, \
  and\ \bibinfo {author} {\bibfnamefont {P.}~\bibnamefont {Singh}},\ }\href
  {\doibase 10.1103/PhysRevD.77.024046} {\bibfield  {journal} {\bibinfo
  {journal} {Phys. Rev. D}\ }\textbf {\bibinfo {volume} {77}},\ \bibinfo
  {pages} {024046} (\bibinfo {year} {2008})}\BibitemShut {NoStop}%
\bibitem [{\citenamefont {Birrell}\ and\ \citenamefont
  {Davies}(1984)}]{Birrell}%
  \BibitemOpen
  \bibfield  {author} {\bibinfo {author} {\bibfnamefont {N.~D.}\ \bibnamefont
  {Birrell}}\ and\ \bibinfo {author} {\bibfnamefont {P.~C.~W.}\ \bibnamefont
  {Davies}},\ }\href@noop {} {\emph {\bibinfo {title} {Quantum Fields in Curved
  Space}}}\ (\bibinfo  {publisher} {Cambridge University Press},\ \bibinfo
  {year} {1984})\BibitemShut {NoStop}%
\bibitem [{\citenamefont {Cortez}\ \emph {et~al.}(2011)\citenamefont {Cortez},
  \citenamefont {Marug\'an}, \citenamefont {Olmedo},\ and\ \citenamefont
  {Velhinho}}]{guille0}%
  \BibitemOpen
  \bibfield  {author} {\bibinfo {author} {\bibfnamefont {J.}~\bibnamefont
  {Cortez}}, \bibinfo {author} {\bibfnamefont {G.~A.~M.}\ \bibnamefont
  {Marug\'an}}, \bibinfo {author} {\bibfnamefont {J.}~\bibnamefont {Olmedo}}, \
  and\ \bibinfo {author} {\bibfnamefont {J.~M.}\ \bibnamefont {Velhinho}},\
  }\href {http://stacks.iop.org/0264-9381/28/i=17/a=172001} {\bibfield
  {journal} {\bibinfo  {journal} {Class. Quant. Grav.}\ }\textbf {\bibinfo
  {volume} {28}},\ \bibinfo {pages} {172001} (\bibinfo {year}
  {2011})}\BibitemShut {NoStop}%
\bibitem [{\citenamefont {Fern\'andez-M\'endez}\ \emph
  {et~al.}(2012)\citenamefont {Fern\'andez-M\'endez}, \citenamefont
  {Mena~Marug\'an}, \citenamefont {Olmedo},\ and\ \citenamefont
  {Velhinho}}]{guille2}%
  \BibitemOpen
  \bibfield  {author} {\bibinfo {author} {\bibfnamefont {M.}~\bibnamefont
  {Fern\'andez-M\'endez}}, \bibinfo {author} {\bibfnamefont {G.~A.}\
  \bibnamefont {Mena~Marug\'an}}, \bibinfo {author} {\bibfnamefont
  {J.}~\bibnamefont {Olmedo}}, \ and\ \bibinfo {author} {\bibfnamefont {J.~M.}\
  \bibnamefont {Velhinho}},\ }\href {\doibase 10.1103/PhysRevD.85.103525}
  {\bibfield  {journal} {\bibinfo  {journal} {Phys. Rev. D}\ }\textbf {\bibinfo
  {volume} {85}},\ \bibinfo {pages} {103525} (\bibinfo {year}
  {2012})}\BibitemShut {NoStop}%
\bibitem [{\citenamefont {Cortez}\ \emph {et~al.}(2012)\citenamefont {Cortez},
  \citenamefont {Mena~Marug\'an}, \citenamefont {Olmedo},\ and\ \citenamefont
  {Velhinho}}]{guille1}%
  \BibitemOpen
  \bibfield  {author} {\bibinfo {author} {\bibfnamefont {J.}~\bibnamefont
  {Cortez}}, \bibinfo {author} {\bibfnamefont {G.~A.}\ \bibnamefont
  {Mena~Marug\'an}}, \bibinfo {author} {\bibfnamefont {J.}~\bibnamefont
  {Olmedo}}, \ and\ \bibinfo {author} {\bibfnamefont {J.~M.}\ \bibnamefont
  {Velhinho}},\ }\href {\doibase 10.1103/PhysRevD.86.104003} {\bibfield
  {journal} {\bibinfo  {journal} {Phys. Rev. D}\ }\textbf {\bibinfo {volume}
  {86}},\ \bibinfo {pages} {104003} (\bibinfo {year} {2012})}\BibitemShut
  {NoStop}%
\bibitem [{\citenamefont
  {Mart\'{\i}n-Mart\'{\i}nez}(2015)}]{Martin-Martinez:2015}%
  \BibitemOpen
  \bibfield  {author} {\bibinfo {author} {\bibfnamefont {E.}~\bibnamefont
  {Mart\'{\i}n-Mart\'{\i}nez}},\ }\href {\doibase 10.1103/PhysRevD.92.104019}
  {\bibfield  {journal} {\bibinfo  {journal} {Phys. Rev. D}\ }\textbf {\bibinfo
  {volume} {92}},\ \bibinfo {pages} {104019} (\bibinfo {year}
  {2015})}\BibitemShut {NoStop}%
\bibitem [{\citenamefont {Pozas-Kerstjens}\ and\ \citenamefont
  {Mart\'{\i}n-Mart\'{\i}nez}(2015)}]{Pozas-Kerstjens:2015}%
  \BibitemOpen
  \bibfield  {author} {\bibinfo {author} {\bibfnamefont {A.}~\bibnamefont
  {Pozas-Kerstjens}}\ and\ \bibinfo {author} {\bibfnamefont {E.}~\bibnamefont
  {Mart\'{\i}n-Mart\'{\i}nez}},\ }\href {\doibase 10.1103/PhysRevD.92.064042}
  {\bibfield  {journal} {\bibinfo  {journal} {Phys. Rev. D}\ }\textbf {\bibinfo
  {volume} {92}},\ \bibinfo {pages} {064042} (\bibinfo {year}
  {2015})}\BibitemShut {NoStop}%
\bibitem [{\citenamefont {Mart\'{i}n-Mart\'{i}nez}\ \emph
  {et~al.}(2013)\citenamefont {Mart\'{i}n-Mart\'{i}nez}, \citenamefont
  {Montero},\ and\ \citenamefont {del Rey}}]{Wavepackets}%
  \BibitemOpen
  \bibfield  {author} {\bibinfo {author} {\bibfnamefont {E.}~\bibnamefont
  {Mart\'{i}n-Mart\'{i}nez}}, \bibinfo {author} {\bibfnamefont
  {M.}~\bibnamefont {Montero}}, \ and\ \bibinfo {author} {\bibfnamefont
  {M.}~\bibnamefont {del Rey}},\ }\href {\doibase 10.1103/PhysRevD.87.064038}
  {\bibfield  {journal} {\bibinfo  {journal} {Phys. Rev. D}\ }\textbf {\bibinfo
  {volume} {87}},\ \bibinfo {pages} {064038} (\bibinfo {year}
  {2013})}\BibitemShut {NoStop}%
\bibitem [{\citenamefont {DeWitt}(1980)}]{DeWitt}%
  \BibitemOpen
  \bibfield  {author} {\bibinfo {author} {\bibfnamefont {B.}~\bibnamefont
  {DeWitt}},\ }\href@noop {} {\emph {\bibinfo {title} {General Relativity; an
  Einstein Centenary Survey}}}\ (\bibinfo  {publisher} {Cambridge University
  Press},\ \bibinfo {year} {1980})\BibitemShut {NoStop}%
\bibitem [{\citenamefont {Alhambra}\ \emph {et~al.}(2014)\citenamefont
  {Alhambra}, \citenamefont {Kempf},\ and\ \citenamefont
  {Mart\'in-Mart\'inez}}]{Alvaro}%
  \BibitemOpen
  \bibfield  {author} {\bibinfo {author} {\bibfnamefont {A.~M.}\ \bibnamefont
  {Alhambra}}, \bibinfo {author} {\bibfnamefont {A.}~\bibnamefont {Kempf}}, \
  and\ \bibinfo {author} {\bibfnamefont {E.}~\bibnamefont
  {Mart\'in-Mart\'inez}},\ }\href {\doibase 10.1103/PhysRevA.89.033835}
  {\bibfield  {journal} {\bibinfo  {journal} {Phys. Rev. A}\ }\textbf {\bibinfo
  {volume} {89}},\ \bibinfo {pages} {033835} (\bibinfo {year}
  {2014})}\BibitemShut {NoStop}%
\bibitem [{\citenamefont {Langlois}(2006)}]{Langlois:2005nf}%
  \BibitemOpen
  \bibfield  {author} {\bibinfo {author} {\bibfnamefont {P.}~\bibnamefont
  {Langlois}},\ }\href {\doibase 10.1016/j.aop.2006.01.013} {\bibfield
  {journal} {\bibinfo  {journal} {Annals Phys.}\ }\textbf {\bibinfo {volume}
  {321}},\ \bibinfo {pages} {2027} (\bibinfo {year} {2006})}\BibitemShut
  {NoStop}%
\bibitem [{\citenamefont {Louko}\ and\ \citenamefont
  {Satz}(2006)}]{Louko:2006zv}%
  \BibitemOpen
  \bibfield  {author} {\bibinfo {author} {\bibfnamefont {J.}~\bibnamefont
  {Louko}}\ and\ \bibinfo {author} {\bibfnamefont {A.}~\bibnamefont {Satz}},\
  }\href {\doibase 10.1088/0264-9381/23/22/015} {\bibfield  {journal} {\bibinfo
   {journal} {Class. Quant. Grav.}\ }\textbf {\bibinfo {volume} {23}},\
  \bibinfo {pages} {6321} (\bibinfo {year} {2006})}\BibitemShut {NoStop}%
\bibitem [{\citenamefont {Louko}\ and\ \citenamefont
  {Satz}(2007)}]{Louko:2006yf}%
  \BibitemOpen
  \bibfield  {author} {\bibinfo {author} {\bibfnamefont {J.}~\bibnamefont
  {Louko}}\ and\ \bibinfo {author} {\bibfnamefont {A.}~\bibnamefont {Satz}},\
  }\href {\doibase 10.1088/1742-6596/68/1/012014} {\bibfield  {journal}
  {\bibinfo  {journal} {J. Phys. Conf. Ser.}\ }\textbf {\bibinfo {volume}
  {68}},\ \bibinfo {pages} {012014} (\bibinfo {year} {2007})}\BibitemShut
  {NoStop}%
\bibitem [{\citenamefont {Abramowitz}\ and\ \citenamefont
  {Stegun}(1972)}]{abramovitz-stegun}%
  \BibitemOpen
  \bibfield  {author} {\bibinfo {author} {\bibfnamefont {M.}~\bibnamefont
  {Abramowitz}}\ and\ \bibinfo {author} {\bibfnamefont {I.~A.}\ \bibnamefont
  {Stegun}},\ }\href@noop {} {\emph {\bibinfo {title} {{Handbook of
  Mathematical Functions with Formulas, Graphs, and Mathematical Tables}}}}\
  (\bibinfo  {publisher} {New York: Dover Publications},\ \bibinfo {year}
  {1972})\BibitemShut {NoStop}%
\bibitem [{\citenamefont {Silverman}(1955)}]{silverman}%
  \BibitemOpen
  \bibfield  {author} {\bibinfo {author} {\bibfnamefont {R.}~\bibnamefont
  {Silverman}},\ }\href {\doibase 10.1109/TIT.1955.1055138} {\bibfield
  {journal} {\bibinfo  {journal} {IEEE Trans. Inf. Theory}\ }\textbf {\bibinfo
  {volume} {1}},\ \bibinfo {pages} {19} (\bibinfo {year} {1955})}\BibitemShut
  {NoStop}%
\bibitem [{\citenamefont {Poisson}\ \emph {et~al.}(2011)\citenamefont
  {Poisson}, \citenamefont {Pound},\ and\ \citenamefont
  {Vega}}]{Poisson:2011nh}%
  \BibitemOpen
  \bibfield  {author} {\bibinfo {author} {\bibfnamefont {E.}~\bibnamefont
  {Poisson}}, \bibinfo {author} {\bibfnamefont {A.}~\bibnamefont {Pound}}, \
  and\ \bibinfo {author} {\bibfnamefont {I.}~\bibnamefont {Vega}},\ }\href
  {\doibase 10.12942/lrr-2011-7} {\bibfield  {journal} {\bibinfo  {journal}
  {Living Rev. Rel.}\ }\textbf {\bibinfo {volume} {14}},\ \bibinfo {pages} {7}
  (\bibinfo {year} {2011})}\BibitemShut {NoStop}%
\bibitem [{\citenamefont {Louko}\ and\ \citenamefont
  {Satz}(2008)}]{Louko:2007mu}%
  \BibitemOpen
  \bibfield  {author} {\bibinfo {author} {\bibfnamefont {J.}~\bibnamefont
  {Louko}}\ and\ \bibinfo {author} {\bibfnamefont {A.}~\bibnamefont {Satz}},\
  }\href {\doibase 10.1088/0264-9381/25/5/055012} {\bibfield  {journal}
  {\bibinfo  {journal} {Class. Quant. Grav.}\ }\textbf {\bibinfo {volume}
  {25}},\ \bibinfo {pages} {055012} (\bibinfo {year} {2008})}\BibitemShut
  {NoStop}%
\end{thebibliography}%

\end{document}